\documentclass[twocolumn]{aastex62}

\newcommand{\arcs}{$^{\prime\prime}$} % Arcseconds
\newcommand{\beq}{\begin{equation}\begin{aligned}}
\newcommand{\eeq}{\end{aligned}\end{equation}}
\newcommand\ddfrac[2]{\frac{\displaystyle #1}{\displaystyle #2}}

\usepackage{scalerel,amssymb}
\usepackage{amsmath}

\usepackage[]{url}
\graphicspath{{./}{figures/}}

\accepted{\today}
\submitjournal{ApJ}

\shorttitle{Dwarf Galaxy SBF Calibration}
\shortauthors{Carlsten et al.}

\begin{document}

\title{Using Surface Brightness Fluctuations to Study Nearby Satellite Galaxy Systems: Calibration and Methodology}

\correspondingauthor{Scott G. Carlsten}
\email{scottgc@princeton.edu}

\author[0000-0002-5382-2898]{Scott G. Carlsten}
\affil{Department of Astrophysical Sciences, 4 Ivy Lane, Princeton University, Princeton, NJ 08544}

\author[0000-0002-1691-8217]{Rachael L. Beaton}
\altaffiliation{Hubble Fellow}
\affiliation{Department of Astrophysical Sciences, 4 Ivy Lane, Princeton University, Princeton, NJ 08544}
\affiliation{The Observatories of the Carnegie Institution for Science, 813 Santa Barbara St., Pasadena, CA~91101\\}

\author[0000-0003-4970-2874]{Johnny P. Greco}
\altaffiliation{NSF Astronomy \& Astrophysics Postdoctoral Fellow}
\affiliation{Center for Cosmology and AstroParticle Physics (CCAPP), The Ohio State University, Columbus, OH 43210, USA}

\author{Jenny E. Greene}
\affil{Department of Astrophysical Sciences, 4 Ivy Lane, Princeton University, Princeton, NJ 08544}

\begin{abstract}
We explore the use of ground-based surface brightness fluctuation (SBF) measurements to constrain distances to nearby dwarf galaxies. Using archival CFHT Megacam imaging data for a sample of 27 nearby dwarfs, we demonstrate that reliable SBF measurements  and distances accurate to 15\% are possible even for very low surface brightness (LSB, $\mu_{i0}>24$ mag/arcsec$^2$) galaxies with modest, $\sim$hour-long exposures with CFHT.  Combining our sample with a recent sample of 7 dwarfs with SBF measured with HST from the literature, we provide the most robust empirical SBF calibration to-date for the blue colors expected for these low mass systems. Our calibration is credible over the color range $0.3\lesssim g-i\lesssim0.8$ mag. It is also the first SBF calibration tied completely to TRGB distances as each galaxy in the sample has a literature TRGB distance. We find that even though the intrinsic scatter in SBF increases for blue galaxies, the rms scatter in the calibration is still $\lesssim0.3$ mag. We verify our measurements by comparing with HST SBF measurements and detailed image simulations. We argue that ground-based SBF is a very useful tool for characterizing dwarf satellite systems and field dwarfs in the nearby, D$\lesssim$20 Mpc universe.
\end{abstract}

\section{Introduction}
The dwarf satellite system of the Milky Way (MW) has long been used to test $\Lambda$CDM predictions for small-scale structure formation. From the many tensions between theory and observations that have arisen over the past two decades \citep[see][for a recent review]{bullock2017}, it is clear that gaining a full understanding of galaxy formation at the low mass end will require a statistical sample of well-characterized satellite systems around galaxies both similar and dissimilar to the MW. While the well-known Missing Satellites Problem and the Too Big To Fail problem appear to be solved for the MW by the addition of baryonic physics \citep[e.g.][]{maccio2010, wetzel2016}, it is possible that the baryonic prescriptions are over-tuned to reproduce the specific properties of Local Group (LG) dwarfs. Additionally, the Planes of Satellites Problem \citep{pawlowski2018} has recently been highlighted as a small scale problem that is not solved by the inclusion of baryonic physics in cosmological simulations. It is currently unclear how the presence of a thin, co-rotating plane of satellite galaxies around the Milky Way \citep{pawlowskiVPOS}, M31 \citep{ibataGPOA}, and Centaurus A \citep{muller_plane} fits into the $\Lambda$CDM paradigm. Further understanding in all of these problems requires a larger sample of dwarf satellites in a wider variety of systems.

Progress in characterizing dwarf galaxy populations outside of the LG has been started by several groups. From this work, it is clear that the two main roadblocks are: (1) finding very low surface brightness (LSB) objects within the virial radius of nearby galaxies (which can project to several degrees on the sky) and (2) confirming that the discovered objects are actually physically associated with a given host and not background contaminants. Recent follow-up studies have highlighted the importance of this second step. \citet{merritt2016} found with HST follow-up that 4/7 LSB galaxies discovered with the Dragonfly telescope array around M101 ($D$=7 Mpc) were, in fact, background galaxies likely associated with the more distant NGC 5485 group. Similarly, HST follow-up by \citet{cohen2018} found that 4/5 LSB galaxies around NGC 4258 were background. Clearly, characterizing the number and properties of dwarfs outside the LG requires an accurate grasp of the distance to these LSB objects.

Currently used methods to verify group membership include: (1) extensive spectroscopic follow-up to all candidate galaxies \citep[e.g.][]{geha2017,spencer2014}, (2) selecting satellites on size and surface brightness cuts that mimic the LG dwarf population \citep[e.g.][]{tanaka2018, ngc3175, muller101, mullerLeo, mullerCenA, bennet2017}, (3) HST follow-up for tip of the red giant branch (TRGB) or surface brightness fluctuation (SBF) distances \cite[e.g.][]{danieli101, cohen2018,df2}, (4) deep ground-based TRGB distances for nearby systems \citep[e.g.][]{mullerTRGB, madcash, smercina2017, smercina2018, crnojevic2014, sand2014, delgado2018}, and (5) statistical subtraction of a background LSB galaxy density \citep[e.g.][]{tanaka2018, park2017, xi2018,speller2014}. Each of these methods have significant draw-backs. (1) requires expensive telescope time and is difficult for quenched, very low surface brightness dwarfs. (2) can only determine the distance to roughly a factor of two and will bias any comparison of the dwarf population of other groups to that of the LG. (3) requires expensive telescope time since generally only one candidate can be imaged at a time due to HST's small field of view. (4) is limited to nearby ($D\lesssim4$ Mpc) systems where the virial volume takes up a large solid angle in the sky and, therefore, is difficult to completely survey. Finally, (5) is a successful strategy for statistical results like luminosity functions (LFs) but is not suitable for measurements requiring specific properties of member systems, such as determining the detailed structure and properties of the satellite system. 

An alternative, promising possibility to efficiently determine distances to LSB dwarfs is ground-based SBF measurements. SBF essentially entails measuring how resolved a galaxy is, which depends on the distance and dominant stellar population in the galaxy. The dependence of SBF on the age and metallicity of the galaxy's dominant stellar population is generally accounted for in a color-dependent calibration formula. SBF is appealing because distances can be measured from the same images in which the LSB galaxies are discovered and ground-based SBF is possible to at least twice the distance of ground-based TRGB, since fully resolved stars are not needed. Ground-based SBF has had a long history \citep[e.g.][]{tonry1988, tonry2001, cantiello2018} and was  pointed out early to have potential for determining the distance to dwarf spheroidal galaxies in the Local Volume. Work by multiple groups \citep[e.g.][]{mieske_sims, mieske_calib, mieske_fornax, jerjen_sculpt, jerjen_cenA, jerjen_field, jerjen_fornax, jerjen_virgo, jerjen_field2, jerjen_sbfcode} derived calibrations, measured the distance to Local Volume field dwarfs, and used SBF to determine group membership of dEs. Their calibrations were either based on theoretical stellar population synthesis (SPS) models or assumed group membership of dwarfs in Fornax or Virgo. These calibrations were very uncertain and, in the case of the theoretical calibration of \citet{jerjen_field}, did not agree with later SPS models. Since these early works, ground-based SBF has not been used to determine distances and group membership for LSB dwarfs. Instead, the emphasis has been on measuring SBF with HST. The time has come to revisit ground-based SBF given that in several cases the data used to discover the LSB dwarfs are of appropriate depth and quality for SBF measurements \citep[e.g.][]{bennet2017, kim2011, johnny2}, as we demonstrate in this paper. 

There are two main obstructions to using SBF in this way: (1) the LSB nature of the dwarfs make the SBF signal weak and careful measurements that consider all sources of contamination to the SBF signal are required and (2) most LSB dwarf satellites are significantly bluer than the cluster ellipticals on which most existing SBF calibrations are based. The uncertainty and spread in the calibrations increases significantly in the blue \citep{mei2005,mieske_calib, blakeslee2009, blakeslee2010, jensen2015,cantiello2018}. The uncertainty in extrapolating existing calibrations towards bluer colors has recently been highlighted in the context of the distance to NGC 1052-DF2 \citep{pvd2018, blakeslee2018, trujillo2018} and the conclusion that the galaxy has an anomalously low mass to light ratio.

In this paper, we address these limitations by measuring the SBF signal for a wide variety of nearby LSB dwarfs ($D< 12$ Mpc) that have archival CFHT MegaCam/MegaPrime imaging \citep{megacam}. These galaxies were chosen to have TRGB distances, so we provide an absolute SBF calibration suitable for the low-mass dwarf galaxies. \citet{jerjen_field} used a few TRGB distances to derive an offset for their otherwise theoretical SBF calibration and \citet{cantiello2007_trgb} used a handful of TRGB distances to determine which (at that time) existing absolute calibration performed the best, but the calibration we derive is the first based solely on TRGB distances. 

This paper is structured as follows: in \S\ref{sec:data} we describe our data selection and sample, in \S\ref{sec:sbf} we describe our SBF measurement methodology, in \S\ref{sec:calib} we present our SBF calibration and compare with stellar population models, and in \S\ref{sec:discuss} we discuss the results in the context of determining the distance to LSB galaxies and conclude.

\section{Data}\label{sec:data}
We start the galaxy selection with the Nearby Galaxy Catalog of \citet{karachentsev}. We restrict our sample to galaxies that have measured TRGB distances in the range $2.5<D<12$ Mpc. The lower bound is to eliminate galaxies that are so resolved that the SBF would be very difficult to measure as a smooth background brightness profile of a galaxy could not be estimated. We supplement this catalog with the recent sample of satellites around NGC 4258 and M96 from \citep{cohen2018} and Do1 from \citet{delgado2018}. 

Each galaxy is searched for in the CFHT MegaCam archive\footnote{\url{http://www.cadc-ccda.hia-iha.nrc-cnrc.gc.ca/en/}}, and only galaxies with both $g$ and $i$ band archival imaging are used. $i$ band is a common choice for measuring SBF, as it is a middle ground between the competing factors that SBF is brighter and seeing is generally better for redder pass-bands \citep[e.g][]{jensen2003, scott_psf}, whereas the instrumental efficiency and the sky foreground become limiting factors in the infrared. We therefore measure SBF in the $i$ band and use the $g-i$ color to parameterize the SBF's dependence on the stellar population. MegaCam has had several generations of $g$ and $i$ filters since first-light in 2003. We accept galaxies imaged in any of the three $i$ filters used (\textsc{I.MP9701}, \textsc{I.MP9702}, and \textsc{I.MP9703}) and either of the two $g$ filters (\textsc{G.MP9401} and \textsc{G.MP9402}). For galaxies that have imaging in more than one $g$ or $i$ filter, we choose the filter with the most exposure time so that each galaxy uses imaging done in only one filter. We show in Appendix \ref{sec:app_filters} that the difference between the filters has an impact smaller than 0.07 mag and, in consequence, we do not attempt to correct for this or try to convert all the $i$ or $g$ band data to a specific $i$ or $g$ filter. All of the different $i$ and $g$ filters are treated equally in the subsequent analysis. All photometry is given in the AB system. Unless stated otherwise, all photometry is corrected for Galactic extinction using the $E(B-V)$ values from \citet{sfd} recalibrated by \citet{sfd2}.

The galaxies are visually inspected and ones with significant substructure that would make SBF measurements difficult are removed. We list the rejected galaxies along with the reason for rejection in Appendix \ref{sec:gal_rejects}. We are left with a sample of 32 galaxies, which are listed in Table \ref{tab:sample}. The sample is inhomogeneous, spanning a range of color, surface brightness, and exposure time.

\begin{deluxetable*}{cccccccccc}

\tablecaption{Main Galaxy Sample}

\tablenum{1}
\label{tab:sample}
\tablehead{\colhead{Name} & \colhead{R.A.} & \colhead{Decl.} & \colhead{Distance} & \colhead{$\mu_{0i}$} & \colhead{$r_{e}$} & \colhead{$g-i$} & \colhead{$M_i$} & \colhead{$t_{\rm exp}$} & \colhead{Multinight} \\ 
\colhead{} & \colhead{} & \colhead{} & \colhead{(Mpc)} & \colhead{(mag arcsec$^{-2}$)} & \colhead{(\arcs)} & \colhead{} & \colhead{(mag)} & \colhead{(sec)} & \colhead{$i$/$g$} } 

\startdata
FM1                  & 9:45:10.0 & +68:45:54 & 3.78     & 24.4$\pm0.1$ & 24.3  & 0.52$\pm0.1$ & -11.8$\pm0.1$ & 660.6    & y/n    \\
UGC 004483            & 8:37:03.0 & +69:46:31 & 3.58     & 22.3$\pm0.1$ & 27.6  & -0.02$\pm0.1$ & -13.3$\pm0.1$ & 600.5    & n/y    \\
KDG 061               & 9:57:02.7 & +68:35:30 & 3.66     & 23.9$\pm0.1$ & 35.3  & 0.39$\pm0.1$ & -12.6$\pm0.1$ & 20003.1  & y/y    \\
BK5N                 & 10:04:40.3 & +68:15:20 & 3.7      & 24.3$\pm0.1$ & 24.4  & 0.52$\pm0.1$ & -11.6$\pm0.1$ & 1101.1   & n/y    \\
LVJ1228+4358         & 12:28:44.9 & +43:58:18 & 4.07     & 25.0$\pm0.1$ & 50.1  & 0.55$\pm0.1$ & -12.1$\pm0.1$ & 1800.8   & n/y    \\
DDO 125$\dagger$               & 12:27:41.9 & +43:29:58 & 2.61     & 22.5$\pm0.1$ & 68.7  & 0.1$\pm0.1$ & -14.5$\pm0.1$ & 1800.8   & n/y    \\
UGCA 365              & 13:36:30.8 & -29:14:11 & 5.42     & 22.1$\pm0.1$ & 27.5  & 0.47$\pm0.1$ & -14.2$\pm0.1$ & 161.1    & n/n    \\
M94\_dw2              & 12:51:04.4 & +41:38:10 & 4.7      & 24.5$\pm0.1$ & 11.0  & 0.6$\pm0.1$ & -9.8$\pm0.1$ & 644.7    & n/n    \\
DDO 044               & 7:34:11.3 & +66:53:10 & 3.21     & 24.2$\pm0.1$ & 46.6  & 0.69$\pm0.1$ & -12.6$\pm0.1$ & 1000.9   & n/n    \\
NGC 4163$\dagger$              & 12:12:08.9 & +36:10:10 & 2.99     & 21.3$\pm0.1$ & 32.9  & 0.56$\pm0.1$ & -14.5$\pm0.1$ & 3601.5   & n/n    \\
NGC 4190$\dagger$              & 12:13:44.6 & +36:37:60 & 2.83     & 20.4$\pm0.1$ & 25.7  & 0.53$\pm0.1$ & -14.7$\pm0.1$ & 3601.5   & n/n    \\
KDG 090               & 12:14:57.9 & +36:13:8 & 2.98     & 23.7$\pm0.1$ & 38.3  & 0.8$\pm0.1$ & -12.6$\pm0.1$ & 3601.5   & n/n    \\
UGC 08508$\dagger$             & 13:30:44.4 & +54:54:36 & 2.67     & 21.8$\pm0.1$ & 30.5  & 0.37$\pm0.1$ & -13.6$\pm0.1$ & 2701.3   & y/n    \\
DDO 190$\dagger$               & 14:24:43.5 & +44:31:33 & 2.83     & 22.1$\pm0.1$ & 32.3  & 0.52$\pm0.1$ & -14.4$\pm0.1$ & 4001.7   & n/n    \\
KKH 98                & 23:45:34.0 & +38:43:04 & 2.58     & 23.2$\pm0.1$ & 21.7  & 0.08$\pm0.1$ & -11.7$\pm0.1$ & 6752.7   & y/y    \\
Do1                  & 1:11:40.4 & +34:36:03 & 3.3      & 25.6$\pm0.1$ & 14.3  & 0.34$\pm0.1$ & -9.1$\pm0.1$ & 3211.6   & y/y    \\
LVJ1218+4655$\dagger$         & 12:18:11.1 & +46:55:01 & 8.28     & 22.1$\pm0.1$ & 19.0  & 0.31$\pm0.1$ & -13.4$\pm0.1$ & 8294.0   & y/y    \\
NGC 4258\_DF6          & 12:19:06.5 & +47:43:51 & 7.3      & 24.4$\pm0.1$ & 8.7   & 0.62$\pm0.1$ & -11.0$\pm0.1$ & 8294.0   & y/y    \\
KDG 101               & 12:19:09.1 & +47:05:23 & 7.28     & 22.7$\pm0.1$ & 26.6  & 0.68$\pm0.1$ & -14.8$\pm0.1$ & 8294.0   & y/y    \\
M101\_DF1             & 14:03:45.0 & +53:56:40 & 6.37     & 24.5$\pm0.1$ & 15.2  & 0.71$\pm0.1$ & -11.0$\pm0.1$ & 4571.6   & y/y    \\
M101\_DF2             & 14:08:37.5 & +54:19:31 & 6.87     & 25.0$\pm0.1$ & 9.3   & 0.69$\pm0.1$ & -10.5$\pm0.1$ & 8612.7   & y/y    \\
M101\_DF3             & 14:03:05.7 & +53:36:56 & 6.52     & 25.7$\pm0.1$ & 29.9  & 0.47$\pm0.1$ & -11.8$\pm0.1$ & 4306.3   & n/y    \\
UGC 9405              & 14:35:24.1 & +57:15:21 & 6.3      & 23.3$\pm0.1$ & 47.3  & 0.58$\pm0.1$ & -15.0$\pm0.1$ & 4306.4   & n/n    \\
M96\_DF9              & 10:46:14.2 & +12:57:38 & 10.0     & 24.0$\pm0.1$ & 8.3   & 0.79$\pm0.1$ & -12.1$\pm0.1$ & 714.6    & y/y    \\
M96\_DF1              & 10:48:13.1 & +11:58:06 & 10.4     & 24.2$\pm0.1$ & 8.6   & 0.57$\pm0.1$ & -11.3$\pm0.1$ & 3930.6   & y/y    \\
M96\_DF8              & 10:46:57.4 & +12:59:55 & 10.2     & 24.0$\pm0.1$ & 23.8  & 0.68$\pm0.1$ & -14.1$\pm0.1$ & 2510.7   & y/y    \\
M96\_DF4              & 10:50:27.2 & +12:21:35 & 10.0     & 24.3$\pm0.1$ & 30.1  & 0.58$\pm0.1$ & -13.9$\pm0.1$ & 952.7    & y/y    \\
M96\_DF5              & 10:49:26.0 & +12:33:10 & 10.8     & 23.6$\pm0.1$ & 28.9  & 0.82$\pm0.1$ & -13.1$\pm0.1$ & 3344.3   & y/y    \\
M96\_DF7              & 10:47:13.5 & +12:48:09 & 10.2     & 25.5$\pm0.1$ & 9.8   & 0.6$\pm0.1$ & -11.2$\pm0.1$ & 2929.9   & y/y    \\
M96\_DF10             & 10:48:36.0 & +13:3:35 & 10.6     & 25.1$\pm0.1$ & 12.5  & 0.77$\pm0.1$ & -11.7$\pm0.1$ & 1915.2   & y/y    \\
M96\_DF6              & 10:46:53.1 & +12:44:34 & 10.2     & 25.5$\pm0.1$ & 23.1  & 0.37$\pm0.1$ & -12.7$\pm0.1$ & 4001.8   & y/y    \\
M96\_DF2              & 10:47:40.6 & +12:02:56 & 10.6     & 25.0$\pm0.1$ & 6.6   & 1.12$\pm0.1$ & -10.7$\pm0.1$ & 4645.3   & y/y    \\
\enddata

\tablecomments{For galaxies indicated with a $\dagger$, the S\'{e}rsic profiles were explicitly fit in the outer regions and the reported surface brightness might not be representative of the actual center. The multinight column indicates whether the data were taken contiguously in one night or over multiple nights. More information on the CFHT data used for each galaxy, including P.I., proposal I.D., and specific filters are given in Table \ref{tab:sample2} in the Appendix. The distances come from \citet{karachentsev} except for the M96 dwarfs and NGC 4258\_DF6 which come from \citet{cohen2018}. The photometry comes from our own measurements.}

\end{deluxetable*}

The default MegaCam pipeline, \texttt{MegaPipe} \citep{megapipe}, is unsuitable for SBF measurements as it performs a very aggressive, local background subtraction that mistakes large LSB galaxies for sky foreground and subtracts them out \citep[e.g.][]{ngvs}. We therefore start with the \texttt{Elixir} \citep{elixir} pre-processed CCD frames and perform the sky subtraction and stacking ourselves. The \texttt{Elixir} pre-processed images have had the instrumental signatures removed and have been flat-fielded. The images are also given a rough astrometric and photometric calibration. We improve upon the astrometric solution by matching sources with SDSS-DR9 \citep{sdss_dr9} sources or USNO-B1 \citep{usnob} for galaxies outside of the SDSS footprint, using the \texttt{Scamp} \citep{scamp} software. The astrometric solutions generally have residuals less than 0.15\arcs rms. The photometric zeropoint for each CCD frame is set by matching sources with SDSS-DR14 \citep{sdss_df14} or Pan-STARRS1-DR1 \citep{panstarrs} for sources outside of the SDSS footprint. SDSS and Pan-STARRS1 magnitudes are converted to the MegaCam photometric system using transformation equations\footnote{Available online \url{http://www.cadc-ccda.hia-iha.nrc-cnrc.gc.ca/en/megapipe/docs/filt.html}} which come from a variety of synthetic and empirical spectral libraries of stars and galaxies. The residuals for the photometry are generally less than 0.05 mag.

The sky subtraction procedure is based on that of the \texttt{Elixir-LSB} pipeline \citep[e.g.][]{ngvs, atlas3d}. The observed background comes from reflections of the sky background in the optics and the flat-fielding process, leading to a radial, `eye'-like pattern on the focal plane. It changes as the sky level changes but is constant over time-scales of a couple hours. We utilize this constancy in time to determine a sky frame for each CCD that is near or covers the galaxy being processed. The data have widely varied dithering patterns but generally there are $>$5 exposures in each filter dithered by 10-20\arcs. These dither sizes are significantly smaller than those used in \citet{ngvs} or \citet{atlas3d}, so large objects ($\gtrsim10$\arcs) would persist after taking the median of the exposures. Therefore we mask sources with \texttt{SExtractor} \citep{sextractor}. We then median combine these exposures for each CCD after scaling by the mode to generate the sky frame for that CCD. This sky frame can then be re-scaled for each individual exposure and subtracted out. The background subtracted images are re-sampled and median combined with a \texttt{Lanczos3} interpolation kernel using \texttt{SWarp} \citep{swarp}. Roughly 1/3 of the galaxies have data that were taken consecutively over a few hours which is the scenario for which the above procedure was designed. The remaining roughly 2/3 of the galaxies have data taken from multiple research groups spanning possibly many years. The background pattern will certainly change on these timescales; however, we find with the simulations described in Appendix \ref{app:sims} that the measurements of the galaxy color and SBF magnitude are not much affected by possible background subtraction errors. 

Once the data are astrometrically and photometrically calibrated and co-added, cutouts around the galaxies are made which are then ready for the SBF measurement.

\section{SBF Measurement}\label{sec:sbf}
In brief, measuring the SBF entails quantifying the brightness fluctuations relative to a smooth background due to Poisson fluctuations in the number of RGB and AGB stars in each resolution element. In line with common definition, we determine the absolute SBF magnitude for each system, which is defined for a stellar population as 

\beq
\bar{M} = -2.5 \log\left(\ddfrac{\sum_i n_i L_i^2}{\sum_i n_i L_i}\right) + \mathrm{z.p.}
\label{eq:sbf_def}
\eeq

\noindent where $n_i$ is the number of stars with luminosity $L_i$ in the stellar population and $\mathrm{z.p.}$ is the zero-point of the photometry. The absolute SBF magnitude depends on the age and metallicity of the dominant stellar population of a galaxy. The goal of the current project is to relate the absolute SBF magnitude with the color of the stellar population, which will account for its dependence on age and metallicity. 

In this section, we describe the major steps in quantifying the SBF. Many of these are standard techniques, though some are particular to our application of SBF to very LSB galaxies.

\subsection{S\'{e}rsic Fits and Masking}
Since a model for the smooth profile of the galaxy is required, the first step in measuring the SBF is fitting a S\'{e}rsic profile to each galaxy. We find that a non-parametric fit to the smooth profile using, for example, elliptical isophotes \citep{ellipse} was not possible for many of the faint dwarf galaxies analyzed here and similarly will not be possible for many faint satellite galaxies that are a primary motivation for the current work. For consistency, we therefore use single S\'{e}rsic profiles as models for the smooth galaxy background profile for all galaxies in our sample. \citet{cohen2018} use this approach as well. As described above, galaxies that display strongly non-axisymmetric profiles or otherwise visibly deviate from a S\'{e}rsic profile are not included in the analysis. As discussed more below, many galaxies will still have subtle deviations from S\'{e}rsic profiles and this will contribute to the scatter of the calibration. For some of the included galaxies, the central regions had complicated clumpy structure whereas the outer regions were well described by a smooth S\'{e}rsic. For these galaxies, the central surface brightness given in Table \ref{tab:sample} comes from fitting the outer regions while masking the inner regions and might not be representative of the actual center. \texttt{Imfit} \citep{imfit} is used to do the fitting. Because it is generally deeper, the $g$-band image is always fit first and then only the normalization of the S\'{e}rsic profile is allowed to vary when fitting the $i$-band image. Due to worse seeing and less SBF, the $g$-band images are also generally less resolved so a smooth S\'{e}rsic is easier to fit.  

Once a S\'{e}rsic profile is fit to each galaxy, it is subtracted out and point sources in the image are detected and masked using \texttt{sep}\footnote{\url{https://github.com/kbarbary/sep}} \citep{sep}, a Python implementation for SExtractor \citep{sextractor}. The masking is to remove contributions to the brightness fluctuations from foreground stars, star clusters in the galaxy analyzed, and background galaxies. Choosing the masking threshold requires some care, as too high of a threshold will leave many sources that bias the fluctuation measurement high and too low of a mask might actually mask some of the brightest SBF. Instead of choosing a fixed threshold defined in terms of standard deviations above the background, we set the mask at a certain absolute magnitude level at the distance of the galaxy. Globular clusters will have $M_{i}\sim-8$ mag while the brightest RGB stars will have $M_{i}\sim-4$ mag. Therefore, there is much room between those magnitudes to set a masking threshold that masks GCs while not masking SBF. Foreground stars and background galaxies are less of a concern since the fluctuation signal in a background control field is subtracted out, as described below. For all the galaxies we mask the images down to sources that would have absolute magnitudes in the range -4 to -6. The specific threshold in that range is chosen on a per-galaxy basis to ensure that bright objects that visually do not look like SBF within the galaxy are masked. These magnitudes correspond to thresholds of 2 to 10 standard deviations above the background, depending on the depth of the data for a particular galaxy. The same magnitude threshold is used for each of the $g$ and $i$ bands and the detection masks are merged between both the filters.

\subsection{Fluctuation Measurement}
After the galaxy has a S\'{e}rsic fit and is masked, we normalize the image by dividing by the square root of the galaxy model. Then we measure the SBF variance in the usual way by Fourier transforming the image and calculating the (1D) power spectrum. This power spectrum is fit by a combination of the power spectrum of the PSF model convolved with the power spectrum of the mask and a constant, representing the power spectrum of (white) photometric noise. Due to the warping in the stacking procedure, the photometric noise is not actually white but using a \texttt{Lanczos3} interpolation kernel will minimize the correlations in the noise \citep{cantiello2005}. We model the PSF at the location of each galaxy using the images of nearby, unsaturated stars. The area of the galaxy used in the fit depends on the galaxy. For most of the LSB galaxies with $\mu_{i0}>23$ mag/arcsec$^2$, the galaxy is fit out to the radius where it falls below 0.4 times the central brightness of the galaxy. This was found to roughly maximize the S/N of the SBF measurement for these faint galaxies. For the galaxies with non-S\'{e}rsic centers, an annulus was used that avoided the central region. The fitted regions had radii anywhere between $50$ to $250$ pixels (10-50\arcs), depending on the size of the galaxy.

\begin{figure*}
\includegraphics[width=\textwidth]{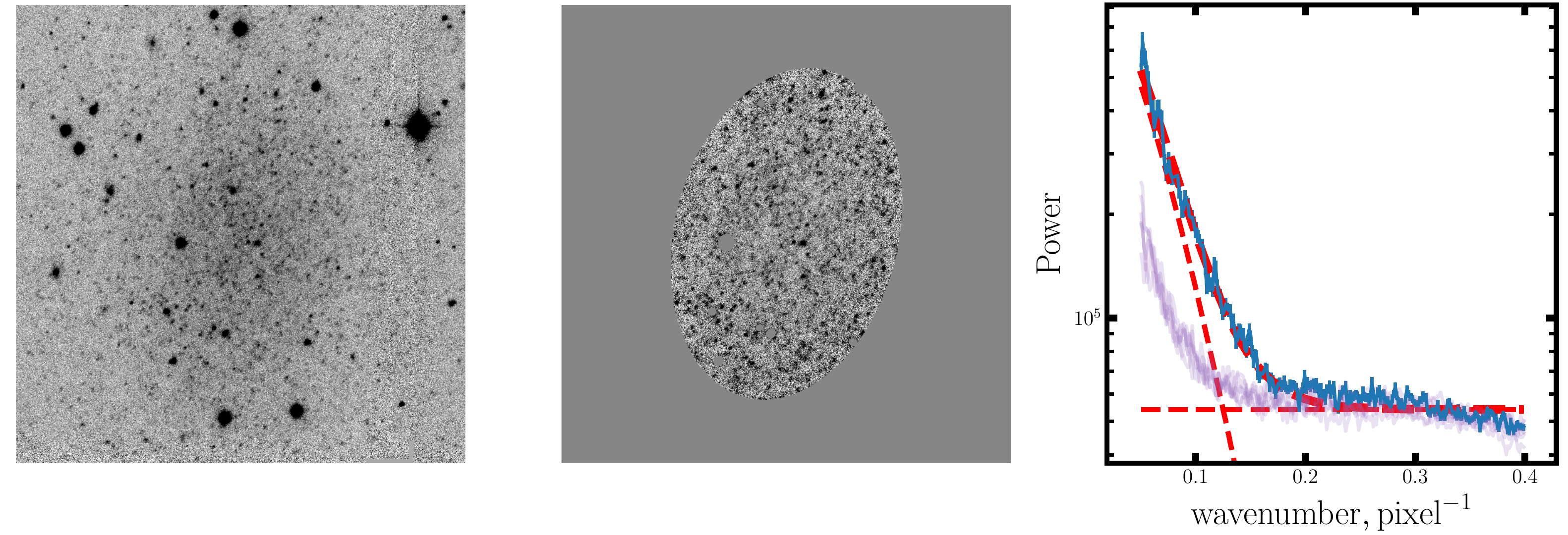}
\caption{Example of the SBF measurement process for the galaxy DDO044. The left panel shows the $i$-band image of the galaxy. The center panel shows the galaxy once the smooth background has been subtracted out, the image divided by the square root of the smooth profile, and the image masked for contaminating sources. The right panel shows the azimuthally averaged power spectrum (in blue) and the best-fitting combination of a white noise component and PSF component to the variance (dashed red). The power spectrum starts to drop at high wavenumber due to the correlated noise present in the images. The purple lines show the power spectra measured in the background fields.}
\label{fig:example_sbf}
\end{figure*}

Unlike the usual approach \citep[e.g.][]{tonry2001, blakeslee2009, cantiello2018}, we do not attempt to model the contribution to the fluctuation signal from undetected (and, hence, unmasked) globular clusters and background galaxies. Instead, we measure the fluctuation signal in nearby fields and subtract this residual level from the signal measured from the galaxy. This approach will remove the residual signal due to unmasked background galaxies and foreground MW stars but will not correct for GCs associated with the galaxy. Since our galaxies are primarily very low mass dwarfs, they are not expected to have many GCs \citep{gc_halo}. The existing ones are bright enough to be easily detected and masked and so GCs and other star clusters are not a concern for these galaxies. This background field approach has the benefit of also accounting for the residual variance due to the correlated noise in the resampling process.

\subsection{Uncertainty in SBF Measurement}

The uncertainty in the SBF measurement comes from two main sources: (1) the stochasticity in the residual signal from unmasked sources and (2) the uncertainty in the actual power spectrum fit for the galaxy. In this section, we describe how we include both sources in our final SBF measurement uncertainty. 

We are in a different regime than most SBF studies \citep[e.g.][]{tonry2001, blakeslee2009, cantiello2018} because our galaxies are much smaller than the big ellipticals usually studied. Therefore, the randomness in the residual signal from unmasked sources contributes significantly to the error in the SBF measurement. To include this in our estimate of the uncertainty of the SBF signal, we measure the SBF in a grid of nearby background fields. The background fields are masked and normalized just like the galaxy (and have the same exposure time and depth). We do a single iteration of 5$\sigma$ clipping of these fields to remove fields that are affected by incompletely masked saturated stars or other artifacts. Then we use the median signal in these fields as an estimate of the residual signal and the standard deviation of the signal in these fields as an estimate of the added uncertainty due to the stochasticity of the residual signal. The residual signal is subtracted from the SBF signal measured from the galaxy and the uncertainty in this residual is added in quadrature to the uncertainty of the SBF measurement from the galaxy (described in the following paragraph). We find that for the majority of the faint ($\mu_0\gtrsim23$ mag/arcsec$^2$) galaxies, this source of uncertainty dominates the overall uncertainty in the SBF measurement.

Following \citet{cohen2018}, we estimate the error in fitting for the SBF magnitude of the galaxy by slightly altering at random the area of the galaxy used and the range of wavenumbers fit in the power spectrum in a Monte Carlo approach. Each galaxy's power spectrum was fit 50 times with each iteration using a slightly different annulus (centered on the fiducial area chosen for each galaxy, as described above) and different lower and upper wavenumbers. The lower wavenumber was chosen in the range 0.01 to 0.1 pixels$^{-1}$ and the upper wavenumber was chosen in the range 0.3 to 0.5 pixels$^{-1}$. We take the median of this distribution to be the measured SBF signal and its standard deviation as the uncertainty. The main steps of this procedure are shown in Figure \ref{fig:example_sbf} for an example galaxy.

We define a rough estimate of the S/N of the SBF detection as:
\beq
\rm{S}/\rm{N} = \frac{P_g - P_{\rm bg}}{\sigma_{\rm bg}}
\eeq
where $P_{\rm g}$ is the variance measured from the galaxy, $P_{\rm bg}$ is the residual variance from the background fields, and $\sigma_{\rm bg}$ is the standard deviation of the residual variance. This definition emphasizes the effect of the stochasticity of the residual variance. Some of the LSB satellites from the \citet{cohen2018} sample in the M96 system were too faint to detect SBF in the CFHT data. We only used the CFHT SBF measurements for sources with S/N $> 2$. Five of the M96 group galaxies had measurable SBF above this threshold. The HST SBF measurements from \citet{cohen2018} are used for the other galaxies in the M96 group, as described below in \S\ref{sec:calib} where we present the calibration.

\begin{figure}
\includegraphics[width=0.46\textwidth]{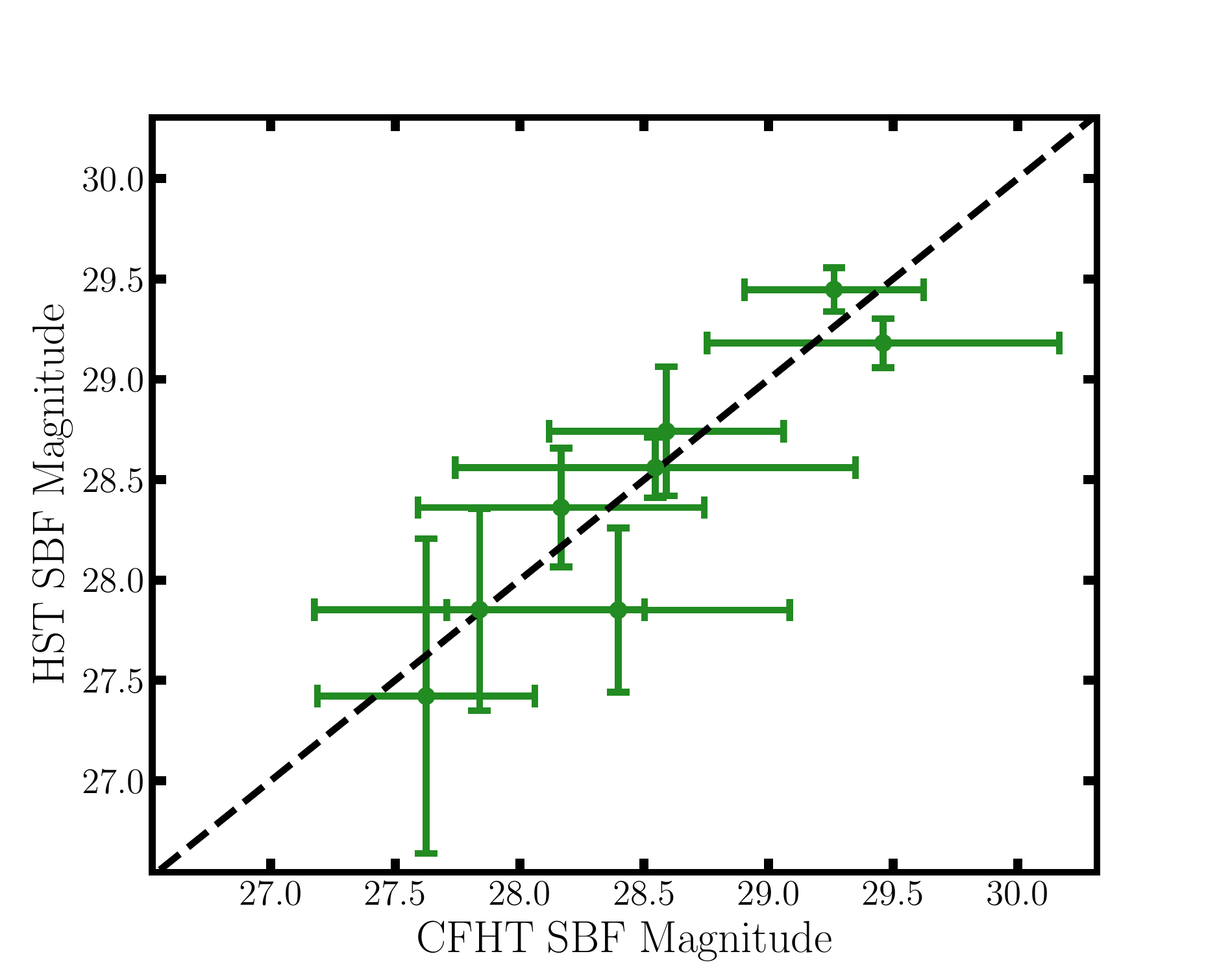}
\caption{Comparison between the SBF (apparent) magnitudes we measure with the CFHT data and those measured from HST data by \citet{cohen2018} for the eight galaxies in common. The dashed line shows a one-to-one correspondence.}
\label{fig:sbf_comparison}
\end{figure}

\subsection{Comparison to HST Measurements}
We have six galaxies in our sample for which we could measure the SBF with the CFHT data and which had HST SBF measurements from \citet{cohen2018}. These galaxies include one from the NGC 4258 group (NGC 4258\_DF6) and five from the M96 group (M96\_DF1, M96\_DF2, M96\_DF4, M96\_DF6, M96\_DF8). We supplement this with two more galaxies from the NGC 4258 area from \citet{cohen2018} (NGC 4258\_DF1 and NGC 4258\_DF2) for which CFHT $i$ band data existed. These two are not in the main galaxy sample because they do not have TRGB distances due to the fact that they are significantly behind NGC 4258 \citep{cohen2018}. As a first verification of our SBF measurements, we compare our measured SBF magnitudes with those measured from the HST data for these galaxies in common. We take the measured (apparent) SBF magnitudes in the HST $I_{814}$-band from \citet{cohen2018} and convert to CFHT $i$-band via
\beq
\bar{m}_i^{\rm CFHT} = \bar{m}_{I_{814}}^{\rm HST} + 0.702 x^2 - 0.852 x + 0.372
\label{eq:cfht_hst_mi}
\eeq
where $x\equiv g_{475} - I_{814}$ is the HST galaxy color from \citet{cohen2018}. We derive this formula from simple stellar population (SSP) predictions of the SBF magnitudes in the two filter systems from the \textsc{MIST} isochrones \citep{mist_models}. We can then directly compare our measured SBF magnitudes with those from \citet{cohen2018} converted to the CFHT filter system. Figure \ref{fig:sbf_comparison} shows this comparison for the eight galaxies in common. All galaxies fall on the 1:1 line within their uncertainties and span roughly two magnitudes in SBF brightness.

\subsection{Image Simulations}
As an additional check for our SBF measurement process, we performed realistic image simulations to see how well we could recover the color and SBF magnitude of an LSB galaxy. The goal of these simulations is three-fold: (1) we explore how well our sky subtraction performs and the accuracy of the measured galaxy colors, (2) we show that the measured SBF magnitude is an unbiased estimator of the true SBF magnitude, and (3) we show that our reported uncertainties are realistic based on the primary sources of error. This last point is a crucial prerequisite to understanding the intrinsic spread in the calibration.

We describe the simulation process in detail in Appendix \ref{app:sims}. We simulate galaxies with the properties of six specific galaxies in our sample: BK5N, NGC 4163, NGC 4258\_DF6, M101\_DF3, KKH98, and M94\_Dw2. These six were chosen as they have a representative range in surface brightness, size, and exposure time (cf. Table \ref{tab:sample}). 

As demonstrated in more detail in the Appendix, we find from the recovered SBF magnitudes that our SBF measurement procedure is an unbiased estimator of the true (input) SBF magnitude. Importantly, we find that the estimated uncertainty in the SBF magnitude is realistic given the spread of the recovered SBF magnitudes. Finally, we find that we recover the colors of the galaxies with an accuracy of $~\sim0.1$ mag which seems to be the limit set by the sky subtraction process. We use this as a representative uncertainty in the galaxy colors and photometry.

\begin{figure*}
\includegraphics[width=0.95\textwidth]{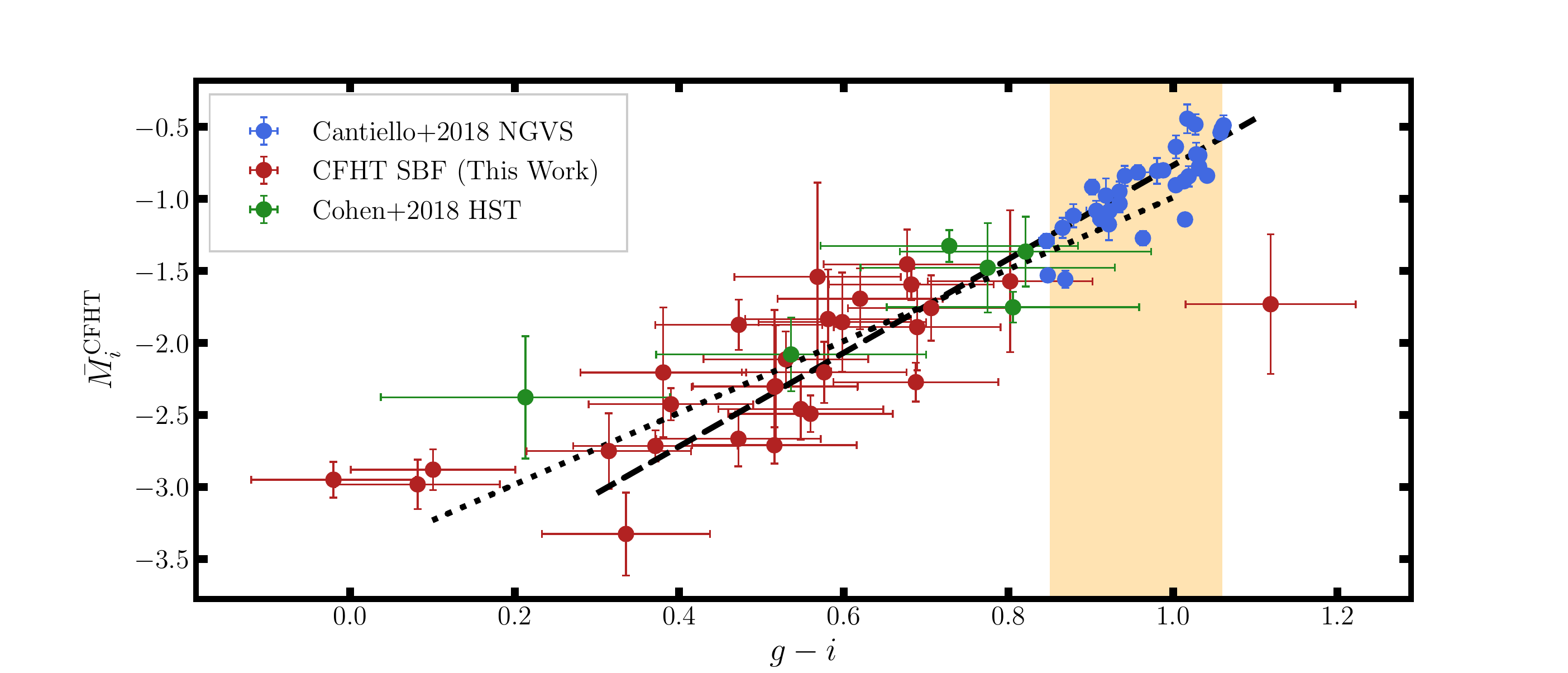}
\caption{Absolute SBF magnitude vs. $g-i$ color for our entire sample, including 6 galaxies with SBF measured by \citet{cohen2018} for which the CFHT imaging was too shallow to measure ground-based SBF. Also shown is the NGVS sample of \citet{cantiello2018}. The orange shaded region is the extent of the calibration from \citet{cantiello2018}, demonstrating the need for bluer calibrations. The dashed black line shows the calibration from \citet{cantiello2018}. The dotted black line shows our best fitting calibration given in the text. The outlier galaxy at $g-i\sim1.1$ is M96\_DF2.}
\label{fig:sbf_calib1}
\end{figure*}

\section{SBF Calibration}\label{sec:calib}
With the SBF magnitudes measured for each galaxy in the sample, we now turn to the primary motivation of this paper: to derive an empirical SBF calibration that is valid for blue, low-mass dwarf galaxies. 

As mentioned above, there were four galaxies in our sample in the M96 system that were too faint to have SBF reliably measured from the CFHT imaging. Instead of completely removing these galaxies from the calibration, we use the HST SBF measurements from \citet{cohen2018}. In addition to these four, we include two other M96 dwarfs from \citet{cohen2018} for which there was no CFHT data (M96\_DF3 and M96\_DF11). We convert the SBF magnitudes from the HST $I_{814}$ filter to the CFHT $i$-band with Equation \ref{eq:cfht_hst_mi}. We convert the HST color from \citet{cohen2018} to CFHT $g-i$ with
\beq
(g-i)^{\rm CFHT} = x - 0.091 x^2 - 0.0062 x - 0.0241
\label{eq:cfht_hst_col}
\eeq
where $x\equiv g_{475} - I_{814}$ is the galaxy color from HST. Like Equation \ref{eq:cfht_hst_mi}, this expression is derived from SSP models from the \textsc{MIST} project \citep{mist_models} and is accurate for a wide range of population age and metallicity.

Using the measured TRGB distances, we convert our measured apparent SBF magnitudes to absolute magnitudes. We include a 0.1 mag error in each of the TRGB distances. This is a characteristic random uncertainty in determining the TRGB (see comparisons of TRGB detection methods in \citet{jang2018} and \citet{beaton2018} for discussion) but does not account for any uncertainty on the TRGB zeropoint.

Figure \ref{fig:sbf_calib1} shows the absolute SBF magnitude vs. $g-i$ color for all galaxies in our sample. The sample includes 28 galaxies which have SBF measured with CFHT data, supplemented by 6 galaxies which have SBF measured by HST. A clear relation where SBF magnitude decreases with decreasing color is seen. Also shown is the Next Generation Virgo Survey (NGVS) sample of \citet{cantiello2018} and range of validity of their calibration. The sample from \citet{cantiello2018} is restricted to those galaxies with HST SBF distances from \citet{blakeslee2009} and are in the Virgo cluster proper (labelled ``V" in their Table 2). The apparent $\bar{m}_i$ magnitudes are converted to absolute SBF magnitudes with the distance modulus to Virgo of 31.09 mag used by \citet{blakeslee2009}. We do not use the individual galaxy distances from \citet{blakeslee2009} because those distances are SBF distances and would mask the intrinsic scatter in the SBF calibration\footnote{Note that this argument assumes that the SBF magnitude in CFHT $i$ band correlates strongly with that in HST $z$ band (the passband that \citet{blakeslee2009} uses). In other words, if a galaxy has anomalously bright SBF in the $i$ band, it would have anomalously bright SBF in the $z$ band as well. This is likely the case as \citet{blakeslee2010} have shown that HST $I$ correlates strongly with HST $z$ band.}. We explore the intrinsic scatter in more detail in \S\ref{sec:calib_scatter}. The fact that almost all of our sample, which is primarily composed of LSB satellites, is bluer than the range of \citet{cantiello2018} highlights the need for a bluer SBF calibration in order to use SBF to study dwarf satellite systems. The very red outlier is M96\_DF2. This galaxy had marginally detected SBF in the CFHT data (S/N~$\gtrsim~2$) and we measure it to be much redder than \citet{cohen2018}. Given its low surface brightness and very small size, it is possible that inaccurate sky subtraction led to a very erroneous color. Due to its separation from the rest of the galaxy sample, we use the color and SBF magnitude measured by \citet{cohen2018} for this galaxy for the remainder of the analysis. For the other galaxies, there appears to be good agreement between the SBF-color relation we measure and that from \citet{cohen2018}. The SBF magnitude does not completely flatten out at blue colors but instead appears to get as bright as $\bar{M}_i\sim-3$ mag at the bluest of colors.

The relation between SBF magnitude and color appears to be roughly linear throughout the entire color range ($0.3\lesssim g-i \lesssim 1.1$~mag). In the presence of significant uncertainties in both SBF magnitude and color, ordinary least squares minimization is inadequate \citep[e.g.][]{hogg2010}. Instead, we adopt the Bayesian algorithm \texttt{LINMIX} developed by \citet{kelly2007}. In short, the algorithm models the independent variable's density as a Gaussian mixture model which allows one to write a simple likelihood function for the observed data, accounting for general covariance in the independent and dependent variables and intrinsic scatter in the relationship. This allows posterior distributions on the parameters of the linear relationship to be modelled through MCMC sampling. This method does not exhibit the bias that generalized $\chi^2$ minimization methods do  \citep[e.g.][]{tremaine2002}, even when accounting for errors in the independent variable. We use the \texttt{python} implementation of \texttt{LINMIX} written by J. Meyers\footnote{\url{https://github.com/jmeyers314/linmix}}. In doing the linear regression, we assume the errors on color and SBF magnitude are uncorrelated and Gaussian and also that the intrinsic scatter in the relationship is Gaussian. We take uniform priors on the slope, normalization, and intrinsic scatter squared.

\begin{figure*}
\includegraphics[width=0.98\textwidth]{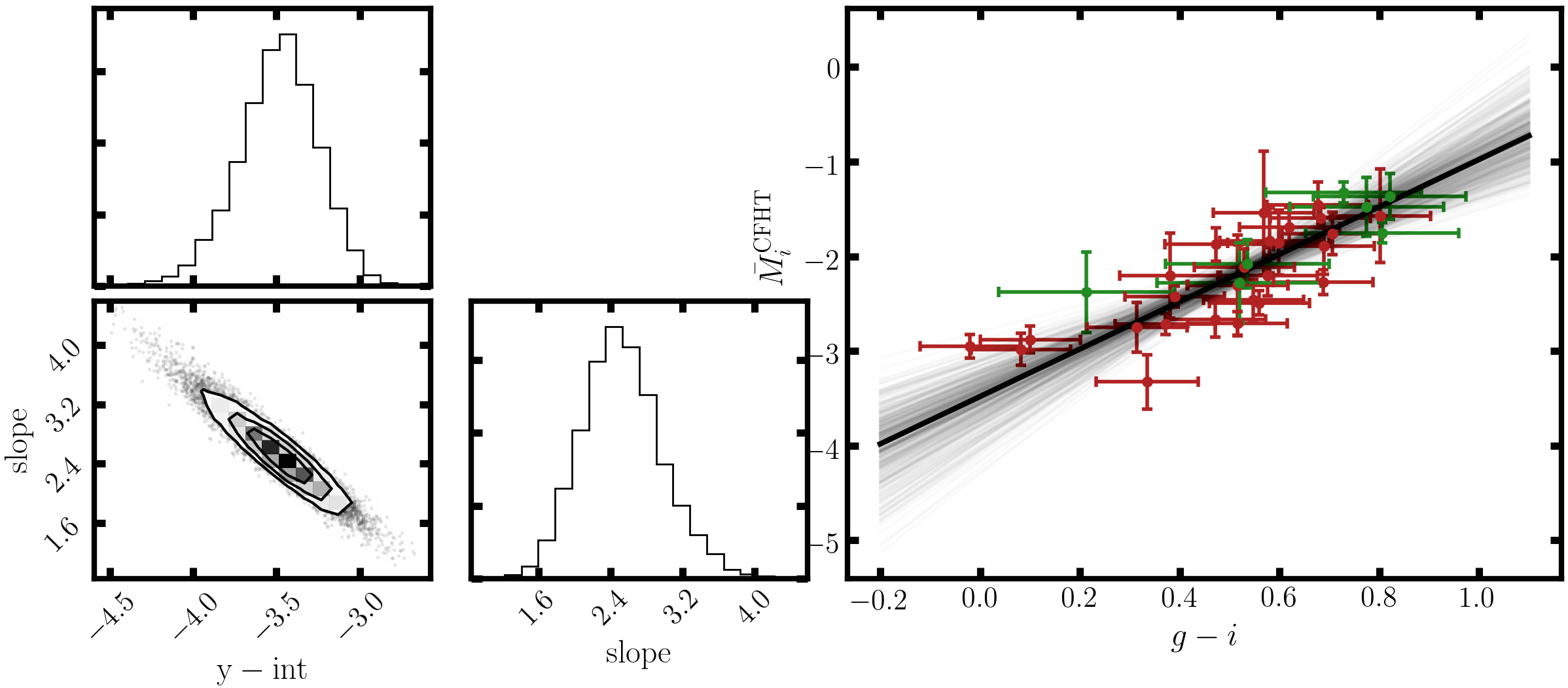}
\caption{Results of linear regression for the LSB sample, including the 7 dwarfs from \citet{cohen2018} (counting M96\_DF2). The left side of the plot shows the posterior distributions from the MCMC samples. The right side shows the SBF magnitude-color relationship for the LSB sample (colors are the same as Figure \ref{fig:sbf_calib1}). The thick black line is the median relation from the linear regression and the thin black lines show random samples from the MCMC chains. \texttt{corner} \citep{corner} was used to visualize the posteriors.} 
\label{fig:calib_mcmc}
\end{figure*}

In Figure \ref{fig:calib_mcmc} we show the results of the linear regression. Marginalized posterior distributions are shown for the $y$-intercept and slope. Taking the median of the posteriors and 1-$\sigma$ uncertainties, we find the linear relationship between SBF magnitude and color as:
\beq
\bar{M}_i = -3.48\pm0.23 + 2.48\pm0.43 (g-i)
\label{eq:calib}
\eeq
We note that the slope is significantly less than the 3.25 found by \citet{cantiello2018} for redder galaxies, indicating that the SBF magnitude-color relation does flatten somewhat at bluer colors. The zero-point of Equation \ref{eq:calib} does not, as written, include any additional uncertainty from the zero-point of the TRGB distances.

\subsection{Scatter of the Calibration}\label{sec:calib_scatter}
An important consideration in the calibration is the intrinsic scatter. Intrinsic scatter in empirical SBF calibrations has been detected before \citep[e.g.][]{blakeslee2009} and is expected theoretically due to the age and metallicity differences in the stellar populations in the galaxies (see \S\ref{sec:calib_ssp} for further discussion of this point). Numerous previous studies have found that the intrinsic spread increases for blue colors \citep{mei2005,mieske_calib, blakeslee2009, blakeslee2010, jensen2015,cantiello2018}. Our TRGB methodology offers the unique capability to measure intrinsic scatter because we are not relying on cluster membership. Empirical calibrations that rely on cluster membership, like those using Virgo cluster galaxies \citep[e.g.][]{blakeslee2009, cantiello2018}, have to disentangle the effects of intrinsic scatter and the significant spread of distances due to the depth of the cluster.

Visually, most of the points in Figure \ref{fig:calib_mcmc} are within the errorbars from the regression line, indicating that the intrinsic scatter is less than the characteristic observational uncertainty. To test this, we compute the generalized reduced $\chi^2$ statistic \citep[e.g.][]{tremaine2002, kelly2007} defined as
\beq
\chi^2_{\rm red} = \frac{1}{N-2}\sum_{i=1}^{N}\frac{(y_i-\alpha-\beta x_i)^2}{\sigma_{yi}^2 + \beta^2\sigma_{xi}^2}
\eeq
where $\alpha$ and $\beta$ are the $y$-intercept and slope of the linear regression, respectively. For our LSB sample, we find that $\chi^2_{\rm red}=0.92$. Since this is less than one, the intrinsic scatter is indeed smaller than the observational scatter and is consistent with being zero. Because the intrinsic scatter is necessarily greater than zero, it is likely that some of the observational uncertainties are over-estimated, particularly the 0.1 mag uncertainty in the color. From the image simulations, we took this as a characteristic limit of the accuracy of the sky subtraction process but it is possibly overly conservative for many of the higher surface brightness galaxies. The inferred intrinsic scatter is clearly very sensitive on the estimated observational uncertainties. Due to the difficulties of estimating the uncertainty introduced by the sky subtraction, we do not attempt to further quantify the intrinsic scatter. Instead, we note that the intrinsic scatter is certainly less than $0.32$ mag which is the residual rms of the regression given in Equation \ref{eq:calib}. This upper-bound is significantly greater than the intrinsic scatter of 0.06 mag estimated for $z$ band SBF for $g-z>1.0$ galaxies by \citet{blakeslee2009}. The rms residual scatter of 0.32 mag (median absolute deviation of 0.21 mag) means that it is possible to measure the distance of these LSB systems to within 15\% accuracy, regardless if the scatter is intrinsic or not.

\subsection{Comparison to Stellar Populations}\label{sec:calib_ssp}
Since SBF probes the second moment of luminosity of a stellar population, it provides somewhat different information than integrated photometry and can be very useful in studying the stellar populations and population gradients in galaxies \citep[e.g.][]{cantiello2007_trgb, cantiello2007, sbf_spectrum}. However, our main goal in this paper is to provide an empirical SBF calibration for LSB dwarfs to study the satellite systems of nearby massive galaxies and an in depth exploration of the stellar populations revealed by SBF is out of our scope. With that said, it is still interesting to compare our SBF magnitude-color relation with that expected from SPS models.

\begin{figure*}
\includegraphics[width=0.95\textwidth]{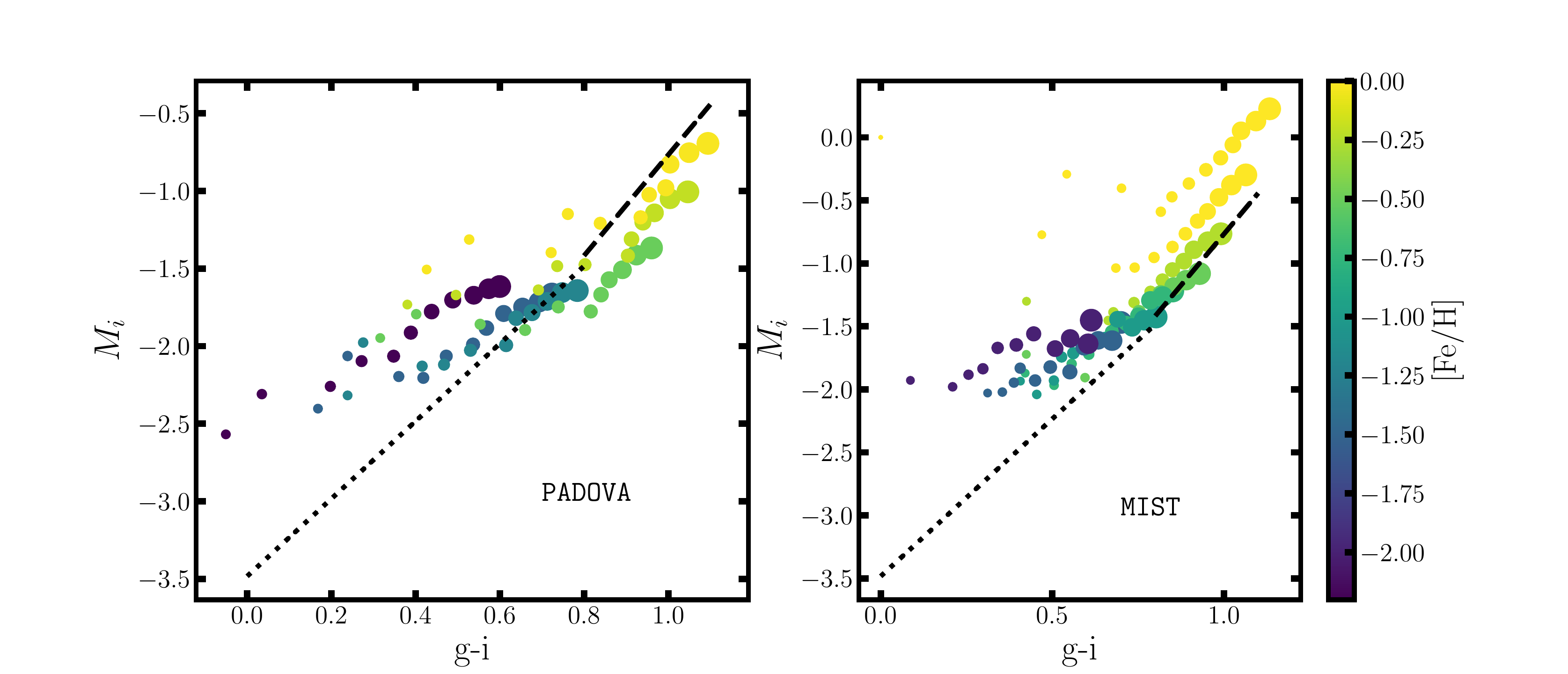}
\caption{{Predicted SBF magnitudes and colors for \texttt{PARSEC} and \textsc{MIST} SSP models in the CFHT filter system. Metallicities in the range $-2<\mathrm{[Fe/H]}<0$ are shown and ages in the range $1<$ age $<10$ Gyr for both sets of models. Points are colored by metallicity and the size reflects the age, with biggest points being the oldest. The dashed line is the best fit calibration of \citet{cantiello2018}. The dotted line shows the regression for our LSB galaxy sample. }}
\label{fig:padova}
\end{figure*}

In Figure \ref{fig:padova} we plot the predicted SBF magnitude-color relation from \textsc{MIST} isochrones \citep{mist_models} and \texttt{PARSEC} isochrones \citep{parsec} with the TP-AGB prescription of \citet{parsec2} for a range in metallicity and age. Younger and more metal poor populations are bluer and have brighter SBF magnitudes, as expected. Also shown are the linear regressions for both the LSB galaxy sample and the sample of \citet{cantiello2018}. Qualitatively the models do reproduce some features of the observations. There is an apparent break in the slope in the models around $g-i\sim0.8$ mag which is also seen in the observations. %Additionally, the scatter in the \texttt{PARSEC} models increases around $g-i\sim0.8$ which agrees with our results in \S\ref{sec:calib_scatter}. For colors redder than $g-i\sim0.8$, decreasing age or metallicity moves one in the same direction in the $\bar{M}_i$ - color plane but for bluer colors, they start to point in slightly different directions, increasing the scatter. 
However, the observations show a steeper slope in the blue than both sets of models. It is unclear whether this is due to the models or some aspect of the SBF measurement for the very bluest galaxies. Perhaps S\'{e}rsic profiles are inadequate or a significant presence of dust adds extra brightness fluctuations. We reserve a complete comparison with models for future work.

\section{Discussion}\label{sec:discuss}
In this paper, we have explored the use of SBF in determining the distance to LSB dwarf galaxies, particularly satellite dwarf spheroidals. We have described an SBF measurement procedure that is optimized for small, faint dwarfs by considering the contribution to the SBF signal from faint, undetected background sources. We have shown that SBF measurement is possible for systems with central surface brightness $>24$ mag/arcsec$^2$, size $\sim10$ arcsec, and distance out to 10 Mpc with modest ($\sim$ hour) exposures on a 4m telescope. We demonstrated the feasibility of these measurements with realistic image simulations and comparison to existing HST measurements.  

We provide an empirical SBF calibration that is valid for the blue colors expected of LSB dwarfs. With 34 galaxies, this is by far the most reliable SBF calibration to-date for blue galaxies. Many of our galaxy sample are LSB satellite galaxies around nearby massive galaxies which make the calibration uniquely adapted for further studies of satellite systems of nearby galaxies since similar stellar populations are expected in these dwarfs. The calibration is given for CFHT $g$ and $i$ band which are close to many modern Sloan-like filter sets in use. Our calibration is also the only calibration to-date that is tied completely to TRGB distances. From Figure \ref{fig:sbf_calib1}, we see that the zero point of the calibration agrees quite well with the Cepheid determined distance to Virgo, lending confidence in the modern distance ladder. 

The rms scatter of the calibration given in Equation \ref{eq:calib} is 0.32 mag. This corresponds to distance measurements with accuracy around $15\%$. While this is not impressive compared to many modern distance estimators, it is more than adequate for studies of satellite galaxies where there is a strong prior on the distance of the dwarf galaxy. Proximity on the sky and a distance accurate to $15\%$ will be sufficient to determine whether a dwarf is associated with a host or not in most cases. 

Comparing to the predictions of state-of-the-art SSP models, we find modest agreement both in the zero-point of the calibration and the slope. The models predict a shallower slope in the absolute SBF magnitude vs. color relation for bluer colors than seen in the data. It is unclear if this is due to incompleteness in the models or due to something in the SBF measurement process, like S\'{e}rsic profiles being inadequate for the very bluest objects.

In conclusion, in this paper, we have argued that SBF has been under-used in studies of nearby (D$\lesssim20$ Mpc) dwarf galaxies. We have demonstrated its potential and provided a robust calibration for its use. Distances accurate to $\sim15\%$ are possible using ground based data with modest exposure times for even very LSB systems. We note that many of the data sets used to discover LSB dwarfs around nearby massive galaxies are likely of the depth and quality to facilitate SBF distances. In a companion paper (Carlsten et al. submitted), we demonstrate this using the CFHT Legacy Survey data used by \citet{bennet2017} to discover LSB dwarf satellite candidates around M101. We are able to confirm 2 galaxies as bona fide companions and many as background galaxies. SBF will be a useful tool in the modern age of deep, wide-field imaging surveys like the Hyper Suprime-Cam Subaru Strategic Program \citep{hsc} and Large Synoptic Survey Telescope \citep{lsst}, possibly even out to distances of 20 Mpc with very deep ground-based data. While our focus has been on satellite dwarfs, SBF will also be useful in discovering and studying LSB field dwarfs \citep[e.g.][]{johnny1, johnny2, danieli_field}. The current work opens up the door to many different research goals from galaxy formation and evolution to dark matter studies.

\section*{Acknowledgements}
% -- RLB 
Support for this work was provided by NASA through Hubble Fellowship grant \#51386.01 awarded to R.L.B.by the Space Telescope Science Institute, which is operated by the Association of  Universities for Research in Astronomy, Inc., for NASA, under contract NAS 5-26555. J.P.G. is supported by an NSF Astronomy and Astrophysics Postdoctoral Fellowship under award AST-1801921. J.E.G. and S.G.C. are partially supported by the National Science Foundation grant AST-1713828.

Based on observations obtained with MegaPrime/MegaCam, a joint project of CFHT and CEA/IRFU, at the Canada-France-Hawaii Telescope (CFHT) which is operated by the National Research Council (NRC) of Canada, the Institut National des Science de l'Univers of the Centre National de la Recherche Scientifique (CNRS) of France, and the University of Hawaii. This work is based in part on data products produced at Terapix available at the Canadian Astronomy Data Centre as part of the Canada-France-Hawaii Telescope Legacy Survey, a collaborative project of NRC and CNRS.

\bibliographystyle{aasjournal}
\bibliography{calib}

\begin{thebibliography}{}
\expandafter\ifx\csname natexlab\endcsname\relax\def\natexlab#1{#1}\fi
\providecommand{\url}[1]{\href{#1}{#1}}

\bibitem[{{Abolfathi} {et~al.}(2018){Abolfathi}, {Aguado}, {Aguilar}, {Allende
  Prieto}, {Almeida}, {Ananna}, {Anders}, {Anderson}, {Andrews}, {Anguiano}, \&
  et~al.}]{sdss_df14}
{Abolfathi}, B., {Aguado}, D.~S., {Aguilar}, G., {et~al.} 2018, \apjs, 235, 42

\bibitem[{{Ahn} {et~al.}(2012){Ahn}, {Alexandroff}, {Allende Prieto},
  {Anderson}, {Anderton}, {Andrews}, {Aubourg}, {Bailey}, {Balbinot}, {Barnes},
  \& et~al.}]{sdss_dr9}
{Ahn}, C.~P., {Alexandroff}, R., {Allende Prieto}, C., {et~al.} 2012, \apjs,
  203, 21

\bibitem[{{Aihara} {et~al.}(2018){Aihara}, {Arimoto}, {Armstrong}, {Arnouts},
  {Bahcall}, {Bickerton}, {Bosch}, {Bundy}, {Capak}, {Chan}, {Chiba}, {Coupon},
  {Egami}, {Enoki}, {Finet}, {Fujimori}, {Fujimoto}, {Furusawa}, {Furusawa},
  {Goto}, {Goulding}, {Greco}, {Greene}, {Gunn}, {Hamana}, {Harikane},
  {Hashimoto}, {Hattori}, {Hayashi}, {Hayashi}, {He{\l}miniak}, {Higuchi},
  {Hikage}, {Ho}, {Hsieh}, {Huang}, {Huang}, {Ikeda}, {Imanishi}, {Inoue},
  {Iwasawa}, {Iwata}, {Jaelani}, {Jian}, {Kamata}, {Karoji}, {Kashikawa},
  {Katayama}, {Kawanomoto}, {Kayo}, {Koda}, {Koike}, {Kojima}, {Komiyama},
  {Konno}, {Koshida}, {Koyama}, {Kusakabe}, {Leauthaud}, {Lee}, {Lin}, {Lin},
  {Lupton}, {Mandelbaum}, {Matsuoka}, {Medezinski}, {Mineo}, {Miyama},
  {Miyatake}, {Miyazaki}, {Momose}, {More}, {More}, {Moritani}, {Moriya},
  {Morokuma}, {Mukae}, {Murata}, {Murayama}, {Nagao}, {Nakata}, {Niida},
  {Niikura}, {Nishizawa}, {Obuchi}, {Oguri}, {Oishi}, {Okabe}, {Okamoto},
  {Okura}, {Ono}, {Onodera}, {Onoue}, {Osato}, {Ouchi}, {Price}, {Pyo}, {Sako},
  {Sawicki}, {Shibuya}, {Shimasaku}, {Shimono}, {Shirasaki}, {Silverman},
  {Simet}, {Speagle}, {Spergel}, {Strauss}, {Sugahara}, {Sugiyama}, {Suto},
  {Suyu}, {Suzuki}, {Tait}, {Takada}, {Takata}, {Tamura}, {Tanaka}, {Tanaka},
  {Tanaka}, {Tanaka}, {Terai}, {Terashima}, {Toba}, {Tominaga}, {Toshikawa},
  {Turner}, {Uchida}, {Uchiyama}, {Umetsu}, {Uraguchi}, {Urata}, {Usuda},
  {Utsumi}, {Wang}, {Wang}, {Wong}, {Yabe}, {Yamada}, {Yamanoi}, {Yasuda},
  {Yeh}, {Yonehara}, \& {Yuma}}]{hsc}
{Aihara}, H., {Arimoto}, N., {Armstrong}, R., {et~al.} 2018, Publications of
  the Astronomical Society of Japan, 70, S4

\bibitem[{{Barbary}(2016)}]{sep}
{Barbary}, K. 2016, JOSS, 1, 58

\bibitem[{{Beaton} {et~al.}(2018){Beaton}, {Bono}, {Braga}, {Dall'Ora},
  {Fiorentino}, {Jang}, {Mart{\'{\i}}nez-V{\'a}zquez}, {Matsunaga}, {Monelli},
  {Neeley}, \& {Salaris}}]{beaton2018}
{Beaton}, R.~L., {Bono}, G., {Braga}, V.~F., {et~al.} 2018, \ssr, 214, 113

\bibitem[{{Bennet} {et~al.}(2017){Bennet}, {Sand}, {Crnojevi{\'c}}, {Spekkens},
  {Zaritsky}, \& {Karunakaran}}]{bennet2017}
{Bennet}, P., {Sand}, D.~J., {Crnojevi{\'c}}, D., {et~al.} 2017, \apj, 850, 109

\bibitem[{{Bertin}(2006)}]{scamp}
{Bertin}, E. 2006, in Astronomical Society of the Pacific Conference Series,
  Vol. 351, Astronomical Data Analysis Software and Systems XV, ed.
  C.~{Gabriel}, C.~{Arviset}, D.~{Ponz}, \& S.~{Enrique}, 112

\bibitem[{{Bertin}(2010)}]{swarp}
{Bertin}, E. 2010, {SWarp: Resampling and Co-adding FITS Images Together},
  Astrophysics Source Code Library, , , ascl:1010.068

\bibitem[{{Bertin} \& {Arnouts}(1996)}]{sextractor}
{Bertin}, E., \& {Arnouts}, S. 1996, \aaps, 117, 393

\bibitem[{{Blakeslee} \& {Cantiello}(2018)}]{blakeslee2018}
{Blakeslee}, J.~P., \& {Cantiello}, M. 2018, Research Notes of the American
  Astronomical Society, 2, 146

\bibitem[{{Blakeslee} {et~al.}(2009){Blakeslee}, {Jord{\'a}n}, {Mei},
  {C{\^o}t{\'e}}, {Ferrarese}, {Infante}, {Peng}, {Tonry}, \&
  {West}}]{blakeslee2009}
{Blakeslee}, J.~P., {Jord{\'a}n}, A., {Mei}, S., {et~al.} 2009, \apj, 694, 556

\bibitem[{{Blakeslee} {et~al.}(2010){Blakeslee}, {Cantiello}, {Mei},
  {C{\^o}t{\'e}}, {Barber DeGraaff}, {Ferrarese}, {Jord{\'a}n}, {Peng},
  {Tonry}, \& {Worthey}}]{blakeslee2010}
{Blakeslee}, J.~P., {Cantiello}, M., {Mei}, S., {et~al.} 2010, \apj, 724, 657

\bibitem[{{Boulade} {et~al.}(2003){Boulade}, {Charlot}, {Abbon}, {Aune},
  {Borgeaud}, {Carton}, {Carty}, {Da Costa}, {Deschamps}, {Desforge},
  {Eppell{\'e}}, {Gallais}, {Gosset}, {Granelli}, {Gros}, {de Kat}, {Loiseau},
  {Ritou}, {Rouss{\'e}}, {Starzynski}, {Vignal}, \& {Vigroux}}]{megacam}
{Boulade}, O., {Charlot}, X., {Abbon}, P., {et~al.} 2003, in \procspie, Vol.
  4841, Instrument Design and Performance for Optical/Infrared Ground-based
  Telescopes, ed. M.~{Iye} \& A.~F.~M. {Moorwood}, 72--81

\bibitem[{{Bressan} {et~al.}(2012){Bressan}, {Marigo}, {Girardi}, {Salasnich},
  {Dal Cero}, {Rubele}, \& {Nanni}}]{parsec}
{Bressan}, A., {Marigo}, P., {Girardi}, L., {et~al.} 2012, \mnras, 427, 127

\bibitem[{{Bullock} \& {Boylan-Kolchin}(2017)}]{bullock2017}
{Bullock}, J.~S., \& {Boylan-Kolchin}, M. 2017, \araa, 55, 343

\bibitem[{{Cantiello} {et~al.}(2007{\natexlab{a}}){Cantiello}, {Blakeslee},
  {Raimondo}, {Brocato}, \& {Capaccioli}}]{cantiello2007_trgb}
{Cantiello}, M., {Blakeslee}, J., {Raimondo}, G., {Brocato}, E., \&
  {Capaccioli}, M. 2007{\natexlab{a}}, \apj, 668, 130

\bibitem[{{Cantiello} {et~al.}(2005){Cantiello}, {Blakeslee}, {Raimondo},
  {Mei}, {Brocato}, \& {Capaccioli}}]{cantiello2005}
{Cantiello}, M., {Blakeslee}, J.~P., {Raimondo}, G., {et~al.} 2005, \apj, 634,
  239

\bibitem[{{Cantiello} {et~al.}(2007{\natexlab{b}}){Cantiello}, {Raimondo},
  {Blakeslee}, {Brocato}, \& {Capaccioli}}]{cantiello2007}
{Cantiello}, M., {Raimondo}, G., {Blakeslee}, J.~P., {Brocato}, E., \&
  {Capaccioli}, M. 2007{\natexlab{b}}, \apj, 662, 940

\bibitem[{{Cantiello} {et~al.}(2018){Cantiello}, {Blakeslee}, {Ferrarese},
  {C{\^o}t{\'e}}, {Roediger}, {Raimondo}, {Peng}, {Gwyn}, {Durrell}, \&
  {Cuillandre}}]{cantiello2018}
{Cantiello}, M., {Blakeslee}, J.~P., {Ferrarese}, L., {et~al.} 2018, \apj, 856,
  126

\bibitem[{{Carlin} {et~al.}(2016){Carlin}, {Sand}, {Price}, {Willman},
  {Karunakaran}, {Spekkens}, {Bell}, {Brodie}, {Crnojevi{\'c}}, {Forbes},
  {Hargis}, {Kirby}, {Lupton}, {Peter}, {Romanowsky}, \& {Strader}}]{madcash}
{Carlin}, J.~L., {Sand}, D.~J., {Price}, P., {et~al.} 2016, \apjl, 828, L5

\bibitem[{{Carlsten} {et~al.}(2018){Carlsten}, {Strauss}, {Lupton}, {Meyers},
  \& {Miyazaki}}]{scott_psf}
{Carlsten}, S.~G., {Strauss}, M.~A., {Lupton}, R.~H., {Meyers}, J.~E., \&
  {Miyazaki}, S. 2018, \mnras, 479, 1491

\bibitem[{{Chambers} {et~al.}(2016){Chambers}, {Magnier}, {Metcalfe},
  {Flewelling}, {Huber}, {Waters}, {Denneau}, {Draper}, {Farrow}, {Finkbeiner},
  {Holmberg}, {Koppenhoefer}, {Price}, {Saglia}, {Schlafly}, {Smartt},
  {Sweeney}, {Wainscoat}, {Burgett}, {Grav}, {Heasley}, {Hodapp}, {Jedicke},
  {Kaiser}, {Kudritzki}, {Luppino}, {Lupton}, {Monet}, {Morgan}, {Onaka},
  {Stubbs}, {Tonry}, {Banados}, {Bell}, {Bender}, {Bernard}, {Botticella},
  {Casertano}, {Chastel}, {Chen}, {Chen}, {Cole}, {Deacon}, {Frenk},
  {Fitzsimmons}, {Gezari}, {Goessl}, {Goggia}, {Goldman}, {Grebel}, {Hambly},
  {Hasinger}, {Heavens}, {Heckman}, {Henderson}, {Henning}, {Holman}, {Hopp},
  {Ip}, {Isani}, {Keyes}, {Koekemoer}, {Kotak}, {Long}, {Lucey}, {Liu},
  {Martin}, {McLean}, {Morganson}, {Murphy}, {Nieto-Santisteban}, {Norberg},
  {Peacock}, {Pier}, {Postman}, {Primak}, {Rae}, {Rest}, {Riess}, {Riffeser},
  {Rix}, {Roser}, {Schilbach}, {Schultz}, {Scolnic}, {Szalay}, {Seitz},
  {Shiao}, {Small}, {Smith}, {Soderblom}, {Taylor}, {Thakar}, {Thiel},
  {Thilker}, {Urata}, {Valenti}, {Walter}, {Watters}, {Werner}, {White},
  {Wood-Vasey}, \& {Wyse}}]{panstarrs}
{Chambers}, K.~C., {Magnier}, E.~A., {Metcalfe}, N., {et~al.} 2016, ArXiv
  e-prints, arXiv:1612.05560

\bibitem[{{Choi} {et~al.}(2016){Choi}, {Dotter}, {Conroy}, {Cantiello},
  {Paxton}, \& {Johnson}}]{mist_models}
{Choi}, J., {Dotter}, A., {Conroy}, C., {et~al.} 2016, \apj, 823, 102

\bibitem[{{Cohen} {et~al.}(2018){Cohen}, {van Dokkum}, {Danieli}, {Romanowsky},
  {Abraham}, {Merritt}, {Zhang}, {Mowla}, {Kruijssen}, {Conroy}, \&
  {Wasserman}}]{cohen2018}
{Cohen}, Y., {van Dokkum}, P., {Danieli}, S., {et~al.} 2018, \apj, 868, 96

\bibitem[{{Crnojevi{\'c}} {et~al.}(2014){Crnojevi{\'c}}, {Sand}, {Caldwell},
  {Guhathakurta}, {McLeod}, {Seth}, {Simon}, {Strader}, \&
  {Toloba}}]{crnojevic2014}
{Crnojevi{\'c}}, D., {Sand}, D.~J., {Caldwell}, N., {et~al.} 2014, \apj, 795,
  L35

\bibitem[{{Danieli} {et~al.}(2018){Danieli}, {van Dokkum}, \&
  {Conroy}}]{danieli_field}
{Danieli}, S., {van Dokkum}, P., \& {Conroy}, C. 2018, \apj, 856, 69

\bibitem[{{Danieli} {et~al.}(2017){Danieli}, {van Dokkum}, {Merritt},
  {Abraham}, {Zhang}, {Karachentsev}, \& {Makarova}}]{danieli101}
{Danieli}, S., {van Dokkum}, P., {Merritt}, A., {et~al.} 2017, \apj, 837, 136

\bibitem[{{Duc} {et~al.}(2015){Duc}, {Cuillandre}, {Karabal}, {Cappellari},
  {Alatalo}, {Blitz}, {Bournaud}, {Bureau}, {Crocker}, {Davies}, {Davis}, {de
  Zeeuw}, {Emsellem}, {Khochfar}, {Krajnovi{\'c}}, {Kuntschner}, {McDermid},
  {Michel-Dansac}, {Morganti}, {Naab}, {Oosterloo}, {Paudel}, {Sarzi}, {Scott},
  {Serra}, {Weijmans}, \& {Young}}]{atlas3d}
{Duc}, P.-A., {Cuillandre}, J.-C., {Karabal}, E., {et~al.} 2015, \mnras, 446,
  120

\bibitem[{{Dunn} \& {Jerjen}(2006)}]{jerjen_sbfcode}
{Dunn}, L.~P., \& {Jerjen}, H. 2006, \aj, 132, 1384

\bibitem[{{Erwin}(2015)}]{imfit}
{Erwin}, P. 2015, \apj, 799, 226

\bibitem[{{Ferrarese} {et~al.}(2012){Ferrarese}, {C{\^o}t{\'e}}, {Cuillandre},
  {Gwyn}, {Peng}, {MacArthur}, {Duc}, {Boselli}, {Mei}, {Erben}, {McConnachie},
  {Durrell}, {Mihos}, {Jord{\'a}n}, {Lan{\c c}on}, {Puzia}, {Emsellem},
  {Balogh}, {Blakeslee}, {van Waerbeke}, {Gavazzi}, {Vollmer}, {Kavelaars},
  {Woods}, {Ball}, {Boissier}, {Courteau}, {Ferriere}, {Gavazzi},
  {Hildebrandt}, {Hudelot}, {Huertas-Company}, {Liu}, {McLaughlin}, {Mellier},
  {Milkeraitis}, {Schade}, {Balkowski}, {Bournaud}, {Carlberg}, {Chapman},
  {Hoekstra}, {Peng}, {Sawicki}, {Simard}, {Taylor}, {Tully}, {van Driel},
  {Wilson}, {Burdullis}, {Mahoney}, \& {Manset}}]{ngvs}
{Ferrarese}, L., {C{\^o}t{\'e}}, P., {Cuillandre}, J.-C., {et~al.} 2012, \apjs,
  200, 4

\bibitem[{{Forbes} {et~al.}(2018){Forbes}, {Read}, {Gieles}, \&
  {Collins}}]{gc_halo}
{Forbes}, D.~A., {Read}, J.~I., {Gieles}, M., \& {Collins}, M. L.~M. 2018,
  \mnras, 481, 5592

\bibitem[{{Foreman-Mackey}(2016)}]{corner}
{Foreman-Mackey}, D. 2016, The Journal of Open Source Software, 1,
  doi:10.21105/joss.00024

\bibitem[{{Geha} {et~al.}(2017){Geha}, {Wechsler}, {Mao}, {Tollerud}, {Weiner},
  {Bernstein}, {Hoyle}, {Marchi}, {Marshall}, {Mu{\~n}oz}, \& {Lu}}]{geha2017}
{Geha}, M., {Wechsler}, R.~H., {Mao}, Y.-Y., {et~al.} 2017, \apj, 847, 4

\bibitem[{{Greco} {et~al.}(2018{\natexlab{a}}){Greco}, {Goulding}, {Greene},
  {Strauss}, {Huang}, {Kim}, \& {Komiyama}}]{johnny1}
{Greco}, J.~P., {Goulding}, A.~D., {Greene}, J.~E., {et~al.}
  2018{\natexlab{a}}, \apj, 866, 112

\bibitem[{{Greco} {et~al.}(2018{\natexlab{b}}){Greco}, {Greene}, {Strauss},
  {Macarthur}, {Flowers}, {Goulding}, {Huang}, {Kim}, {Komiyama}, {Leauthaud},
  {Leisman}, {Lupton}, {Sif{\'o}n}, \& {Wang}}]{johnny2}
{Greco}, J.~P., {Greene}, J.~E., {Strauss}, M.~A., {et~al.} 2018{\natexlab{b}},
  \apj, 857, 104

\bibitem[{{Gwyn}(2008)}]{megapipe}
{Gwyn}, S.~D.~J. 2008, \pasp, 120, 212

\bibitem[{{Hogg} {et~al.}(2010){Hogg}, {Bovy}, \& {Lang}}]{hogg2010}
{Hogg}, D.~W., {Bovy}, J., \& {Lang}, D. 2010, arXiv e-prints, arXiv:1008.4686

\bibitem[{{Ibata} {et~al.}(2013){Ibata}, {Lewis}, {Conn}, {Irwin},
  {McConnachie}, {Chapman}, {Collins}, {Fardal}, {Ferguson}, {Ibata}, {Mackey},
  {Martin}, {Navarro}, {Rich}, {Valls-Gabaud}, \& {Widrow}}]{ibataGPOA}
{Ibata}, R.~A., {Lewis}, G.~F., {Conn}, A.~R., {et~al.} 2013, \nat, 493, 62

\bibitem[{{Jang} {et~al.}(2018){Jang}, {Hatt}, {Beaton}, {Lee}, {Freedman},
  {Madore}, {Hoyt}, {Monson}, {Rich}, {Scowcroft}, \& {Seibert}}]{jang2018}
{Jang}, I.~S., {Hatt}, D., {Beaton}, R.~L., {et~al.} 2018, \apj, 852, 60

\bibitem[{{Jedrzejewski}(1987)}]{ellipse}
{Jedrzejewski}, R.~I. 1987, \mnras, 226, 747

\bibitem[{{Jensen} {et~al.}(2015){Jensen}, {Blakeslee}, {Gibson}, {Lee},
  {Cantiello}, {Raimondo}, {Boyer}, \& {Cho}}]{jensen2015}
{Jensen}, J.~B., {Blakeslee}, J.~P., {Gibson}, Z., {et~al.} 2015, \apj, 808, 91

\bibitem[{{Jensen} {et~al.}(2003){Jensen}, {Tonry}, {Barris}, {Thompson},
  {Liu}, {Rieke}, {Ajhar}, \& {Blakeslee}}]{jensen2003}
{Jensen}, J.~B., {Tonry}, J.~L., {Barris}, B.~J., {et~al.} 2003, \apj, 583, 712

\bibitem[{{Jerjen}(2003)}]{jerjen_fornax}
{Jerjen}, H. 2003, \aap, 398, 63

\bibitem[{{Jerjen} {et~al.}(2004){Jerjen}, {Binggeli}, \&
  {Barazza}}]{jerjen_virgo}
{Jerjen}, H., {Binggeli}, B., \& {Barazza}, F.~D. 2004, \aj, 127, 771

\bibitem[{{Jerjen} {et~al.}(1998){Jerjen}, {Freeman}, \&
  {Binggeli}}]{jerjen_sculpt}
{Jerjen}, H., {Freeman}, K.~C., \& {Binggeli}, B. 1998, \aj, 116, 2873

\bibitem[{{Jerjen} {et~al.}(2000){Jerjen}, {Freeman}, \&
  {Binggeli}}]{jerjen_cenA}
---. 2000, \aj, 119, 166

\bibitem[{{Jerjen} {et~al.}(2001){Jerjen}, {Rekola}, {Takalo}, {Coleman}, \&
  {Valtonen}}]{jerjen_field}
{Jerjen}, H., {Rekola}, R., {Takalo}, L., {Coleman}, M., \& {Valtonen}, M.
  2001, \aap, 380, 90

\bibitem[{{Karachentsev} {et~al.}(2013){Karachentsev}, {Makarov}, \&
  {Kaisina}}]{karachentsev}
{Karachentsev}, I.~D., {Makarov}, D.~I., \& {Kaisina}, E.~I. 2013, \aj, 145,
  101

\bibitem[{{Kelly}(2007)}]{kelly2007}
{Kelly}, B.~C. 2007, \apj, 665, 1489

\bibitem[{{Kim} {et~al.}(2011){Kim}, {Kim}, {Hwang}, {Lee}, {Chun}, \&
  {Ann}}]{kim2011}
{Kim}, E., {Kim}, M., {Hwang}, N., {et~al.} 2011, \mnras, 412, 1881

\bibitem[{{Kondapally} {et~al.}(2018){Kondapally}, {Russell}, {Conselice}, \&
  {Penny}}]{ngc3175}
{Kondapally}, R., {Russell}, G.~A., {Conselice}, C.~J., \& {Penny}, S.~J. 2018,
  \mnras, 481, 1759

\bibitem[{{Kroupa}(2001)}]{kroupa_imf}
{Kroupa}, P. 2001, \mnras, 322, 231

\bibitem[{{LSST Science Collaboration} {et~al.}(2009){LSST Science
  Collaboration}, {Abell}, {Allison}, {Anderson}, {Andrew}, {Angel}, {Armus},
  {Arnett}, {Asztalos}, {Axelrod}, \& et~al.}]{lsst}
{LSST Science Collaboration}, {Abell}, P.~A., {Allison}, J., {et~al.} 2009,
  arXiv e-prints, arXiv:0912.0201

\bibitem[{{Macci{\`o}} {et~al.}(2010){Macci{\`o}}, {Kang}, {Fontanot},
  {Somerville}, {Koposov}, \& {Monaco}}]{maccio2010}
{Macci{\`o}}, A.~V., {Kang}, X., {Fontanot}, F., {et~al.} 2010, \mnras, 402,
  1995

\bibitem[{{Magnier} \& {Cuillandre}(2004)}]{elixir}
{Magnier}, E.~A., \& {Cuillandre}, J.-C. 2004, \pasp, 116, 449

\bibitem[{{Marigo} {et~al.}(2017){Marigo}, {Girardi}, {Bressan}, {Rosenfield},
  {Aringer}, {Chen}, {Dussin}, {Nanni}, {Pastorelli}, {Rodrigues}, {Trabucchi},
  {Bladh}, {Dalcanton}, {Groenewegen}, {Montalb{\'a}n}, \& {Wood}}]{parsec2}
{Marigo}, P., {Girardi}, L., {Bressan}, A., {et~al.} 2017, \apj, 835, 77

\bibitem[{{Martinez-Delgado} {et~al.}(2018){Martinez-Delgado}, {Grebel},
  {Javanmardi}, {Boschin}, {Longeard}, {Carballo-Bello}, {Makarov}, {Beasley},
  {Donatiello}, {Haynes}, {Forbes}, \& {Romanowsky}}]{delgado2018}
{Martinez-Delgado}, D., {Grebel}, E.~K., {Javanmardi}, B., {et~al.} 2018, ArXiv
  e-prints, arXiv:1810.04741

\bibitem[{{Mei} {et~al.}(2005){Mei}, {Blakeslee}, {Tonry}, {Jord{\'a}n},
  {Peng}, {C{\^o}t{\'e}}, {Ferrarese}, {West}, {Merritt}, \&
  {Milosavljevi{\'c}}}]{mei2005}
{Mei}, S., {Blakeslee}, J.~P., {Tonry}, J.~L., {et~al.} 2005, \apj, 625, 121

\bibitem[{{Merritt} {et~al.}(2016){Merritt}, {van Dokkum}, {Danieli},
  {Abraham}, {Zhang}, {Karachentsev}, \& {Makarova}}]{merritt2016}
{Merritt}, A., {van Dokkum}, P., {Danieli}, S., {et~al.} 2016, \apj, 833, 168

\bibitem[{{Mieske} {et~al.}(2003){Mieske}, {Hilker}, \&
  {Infante}}]{mieske_sims}
{Mieske}, S., {Hilker}, M., \& {Infante}, L. 2003, \aap, 403, 43

\bibitem[{{Mieske} {et~al.}(2006){Mieske}, {Hilker}, \&
  {Infante}}]{mieske_calib}
---. 2006, \aap, 458, 1013

\bibitem[{{Mieske} {et~al.}(2007){Mieske}, {Hilker}, {Infante}, \& {Mendes de
  Oliveira}}]{mieske_fornax}
{Mieske}, S., {Hilker}, M., {Infante}, L., \& {Mendes de Oliveira}, C. 2007,
  \aap, 463, 503

\bibitem[{{Mitzkus} {et~al.}(2018){Mitzkus}, {Walcher}, {Roth}, {Coelho},
  {Cioni}, {Raimondo}, \& {Rejkuba}}]{sbf_spectrum}
{Mitzkus}, M., {Walcher}, C.~J., {Roth}, M.~M., {et~al.} 2018, \mnras, 480, 629

\bibitem[{{Monet} {et~al.}(2003){Monet}, {Levine}, {Canzian}, {Ables}, {Bird},
  {Dahn}, {Guetter}, {Harris}, {Henden}, {Leggett}, {Levison}, {Luginbuhl},
  {Martini}, {Monet}, {Munn}, {Pier}, {Rhodes}, {Riepe}, {Sell}, {Stone},
  {Vrba}, {Walker}, {Westerhout}, {Brucato}, {Reid}, {Schoening}, {Hartley},
  {Read}, \& {Tritton}}]{usnob}
{Monet}, D.~G., {Levine}, S.~E., {Canzian}, B., {et~al.} 2003, \aj, 125, 984

\bibitem[{{M{\"u}ller} {et~al.}(2017{\natexlab{a}}){M{\"u}ller}, {Jerjen}, \&
  {Binggeli}}]{mullerCenA}
{M{\"u}ller}, O., {Jerjen}, H., \& {Binggeli}, B. 2017{\natexlab{a}}, \aap,
  597, A7

\bibitem[{{M{\"u}ller} {et~al.}(2018{\natexlab{a}}){M{\"u}ller}, {Jerjen}, \&
  {Binggeli}}]{mullerLeo}
---. 2018{\natexlab{a}}, \aap, 615, A105

\bibitem[{{M{\"u}ller} {et~al.}(2018{\natexlab{b}}){M{\"u}ller}, {Pawlowski},
  {Jerjen}, \& {Lelli}}]{muller_plane}
{M{\"u}ller}, O., {Pawlowski}, M.~S., {Jerjen}, H., \& {Lelli}, F.
  2018{\natexlab{b}}, Science, 359, 534

\bibitem[{{M{\"u}ller} {et~al.}(2018{\natexlab{c}}){M{\"u}ller}, {Rejkuba}, \&
  {Jerjen}}]{mullerTRGB}
{M{\"u}ller}, O., {Rejkuba}, M., \& {Jerjen}, H. 2018{\natexlab{c}}, \aap, 615,
  A96

\bibitem[{{M{\"u}ller} {et~al.}(2017{\natexlab{b}}){M{\"u}ller}, {Scalera},
  {Binggeli}, \& {Jerjen}}]{muller101}
{M{\"u}ller}, O., {Scalera}, R., {Binggeli}, B., \& {Jerjen}, H.
  2017{\natexlab{b}}, \aap, 602, A119

\bibitem[{{Park} {et~al.}(2017){Park}, {Moon}, {Zaritsky}, {Pak}, {Lee}, {Kim},
  {Kim}, \& {Cha}}]{park2017}
{Park}, H.~S., {Moon}, D.-S., {Zaritsky}, D., {et~al.} 2017, \apj, 848, 19

\bibitem[{{Pawlowski}(2018)}]{pawlowski2018}
{Pawlowski}, M.~S. 2018, Modern Physics Letters A, 33, 1830004

\bibitem[{{Pawlowski} {et~al.}(2012){Pawlowski}, {Pflamm-Altenburg}, \&
  {Kroupa}}]{pawlowskiVPOS}
{Pawlowski}, M.~S., {Pflamm-Altenburg}, J., \& {Kroupa}, P. 2012, \mnras, 423,
  1109

\bibitem[{{Rekola} {et~al.}(2005){Rekola}, {Jerjen}, \&
  {Flynn}}]{jerjen_field2}
{Rekola}, R., {Jerjen}, H., \& {Flynn}, C. 2005, \aap, 437, 823

\bibitem[{{Sand} {et~al.}(2014){Sand}, {Crnojevi{\'c}}, {Strader}, {Toloba},
  {Simon}, {Caldwell}, {Guhathakurta}, {McLeod}, \& {Seth}}]{sand2014}
{Sand}, D.~J., {Crnojevi{\'c}}, D., {Strader}, J., {et~al.} 2014, \apj, 793, L7

\bibitem[{{Schlafly} \& {Finkbeiner}(2011)}]{sfd2}
{Schlafly}, E.~F., \& {Finkbeiner}, D.~P. 2011, \apj, 737, 103

\bibitem[{{Schlegel} {et~al.}(1998){Schlegel}, {Finkbeiner}, \& {Davis}}]{sfd}
{Schlegel}, D.~J., {Finkbeiner}, D.~P., \& {Davis}, M. 1998, \apj, 500, 525

\bibitem[{{Smercina} {et~al.}(2018){Smercina}, {Bell}, {Price}, {D Souza},
  {Slater}, {Bailin}, {Monachesi}, \& {Nidever}}]{smercina2018}
{Smercina}, A., {Bell}, E.~F., {Price}, P.~A., {et~al.} 2018, \apj, 863, 152

\bibitem[{{Smercina} {et~al.}(2017){Smercina}, {Bell}, {Slater}, {Price},
  {Bailin}, \& {Monachesi}}]{smercina2017}
{Smercina}, A., {Bell}, E.~F., {Slater}, C.~T., {et~al.} 2017, \apjl, 843, L6

\bibitem[{{Speller} \& {Taylor}(2014)}]{speller2014}
{Speller}, R., \& {Taylor}, J.~E. 2014, \apj, 788, 188

\bibitem[{{Spencer} {et~al.}(2014){Spencer}, {Loebman}, \&
  {Yoachim}}]{spencer2014}
{Spencer}, M., {Loebman}, S., \& {Yoachim}, P. 2014, \apj, 788, 146

\bibitem[{{Tanaka} {et~al.}(2018){Tanaka}, {Chiba}, {Hayashi}, {Komiyama},
  {Okamoto}, {Cooper}, {Okamoto}, \& {Spitler}}]{tanaka2018}
{Tanaka}, M., {Chiba}, M., {Hayashi}, K., {et~al.} 2018, \apj, 865, 125

\bibitem[{{Tonry} \& {Schneider}(1988)}]{tonry1988}
{Tonry}, J., \& {Schneider}, D.~P. 1988, \aj, 96, 807

\bibitem[{{Tonry} {et~al.}(2001){Tonry}, {Dressler}, {Blakeslee}, {Ajhar},
  {Fletcher}, {Luppino}, {Metzger}, \& {Moore}}]{tonry2001}
{Tonry}, J.~L., {Dressler}, A., {Blakeslee}, J.~P., {et~al.} 2001, \apj, 546,
  681

\bibitem[{{Tremaine} {et~al.}(2002){Tremaine}, {Gebhardt}, {Bender}, {Bower},
  {Dressler}, {Faber}, {Filippenko}, {Green}, {Grillmair}, {Ho}, {Kormendy},
  {Lauer}, {Magorrian}, {Pinkney}, \& {Richstone}}]{tremaine2002}
{Tremaine}, S., {Gebhardt}, K., {Bender}, R., {et~al.} 2002, \apj, 574, 740

\bibitem[{{Trujillo} {et~al.}(2018){Trujillo}, {Beasley}, {Borlaff},
  {Carrasco}, {Di Cintio}, {Filho}, {Monelli}, {Montes}, {Roman}, {Ruiz-Lara},
  {Sanchez Almeida}, {Valls-Gabaud}, \& {Vazdekis}}]{trujillo2018}
{Trujillo}, I., {Beasley}, M.~A., {Borlaff}, A., {et~al.} 2018, ArXiv e-prints,
  arXiv:1806.10141

\bibitem[{{van Dokkum} {et~al.}(2018{\natexlab{a}}){van Dokkum}, {Danieli},
  {Cohen}, {Romanowsky}, \& {Conroy}}]{pvd2018}
{van Dokkum}, P., {Danieli}, S., {Cohen}, Y., {Romanowsky}, A.~J., \& {Conroy},
  C. 2018{\natexlab{a}}, \apjl, 864, L18

\bibitem[{{van Dokkum} {et~al.}(2018{\natexlab{b}}){van Dokkum}, {Danieli},
  {Cohen}, {Merritt}, {Romanowsky}, {Abraham}, {Brodie}, {Conroy}, {Lokhorst},
  {Mowla}, {O'Sullivan}, \& {Zhang}}]{df2}
{van Dokkum}, P., {Danieli}, S., {Cohen}, Y., {et~al.} 2018{\natexlab{b}},
  \nat, 555, 629

\bibitem[{{Wetzel} {et~al.}(2016){Wetzel}, {Hopkins}, {Kim},
  {Faucher-Gigu{\`e}re}, {Kere{\v s}}, \& {Quataert}}]{wetzel2016}
{Wetzel}, A.~R., {Hopkins}, P.~F., {Kim}, J.-h., {et~al.} 2016, \apjl, 827, L23

\bibitem[{{Xi} {et~al.}(2018){Xi}, {Taylor}, {Massey}, {Rhodes}, {Koekemoer},
  \& {Salvato}}]{xi2018}
{Xi}, C., {Taylor}, J.~E., {Massey}, R.~J., {et~al.} 2018, \mnras, 478, 5336

\end{thebibliography}

\appendix

\section{SBF Simulations}\label{app:sims}
In this section, we use realistic image simulations to further verify our SBF measurements. In the simulations, we use the $i$-band S\'{e}rsic fits for the galaxies procured in the SBF measurement to generate artificial galaxies that we then insert into the CCD frames (before sky subtraction). We use isochrones from the \textsc{MIST} project \citep{mist_models} to give the artificial galaxies realistic SBF. For a given isochrone, we generate a large sample of stars by sampling a Kroupa IMF \citep{kroupa_imf}. Then, for each pixel in the artificial galaxy, we calculate the expected number of stars in that pixel based on the intensity of the S\'{e}rsic profile at that pixel and the average stellar luminosity in the sample of stars from the isochrone\footnote{The simulated galaxies are assumed to be at the same distance as the real ones that they mimic.}. A number is then drawn from a Poisson distribution with mean equal to this expected number of stars. That number of stars are then drawn from the large sample of stars and put into that pixel. This procedure naturally reproduces the SBF magnitude of the population given in Equation \ref{eq:sbf_def} and also will exhibit realistic stochasticity in the SBF magnitude due to partial sampling of the isochrones because of the low stellar mass of many of our galaxies. This procedure also allows us to simultaneously simulate $i$ and $g$ band images of the artificial galaxies. Finally, the galaxies are convolved with a model for the PSF before being inserted into the CCD frames.

\begin{figure*}
\includegraphics[width=\textwidth]{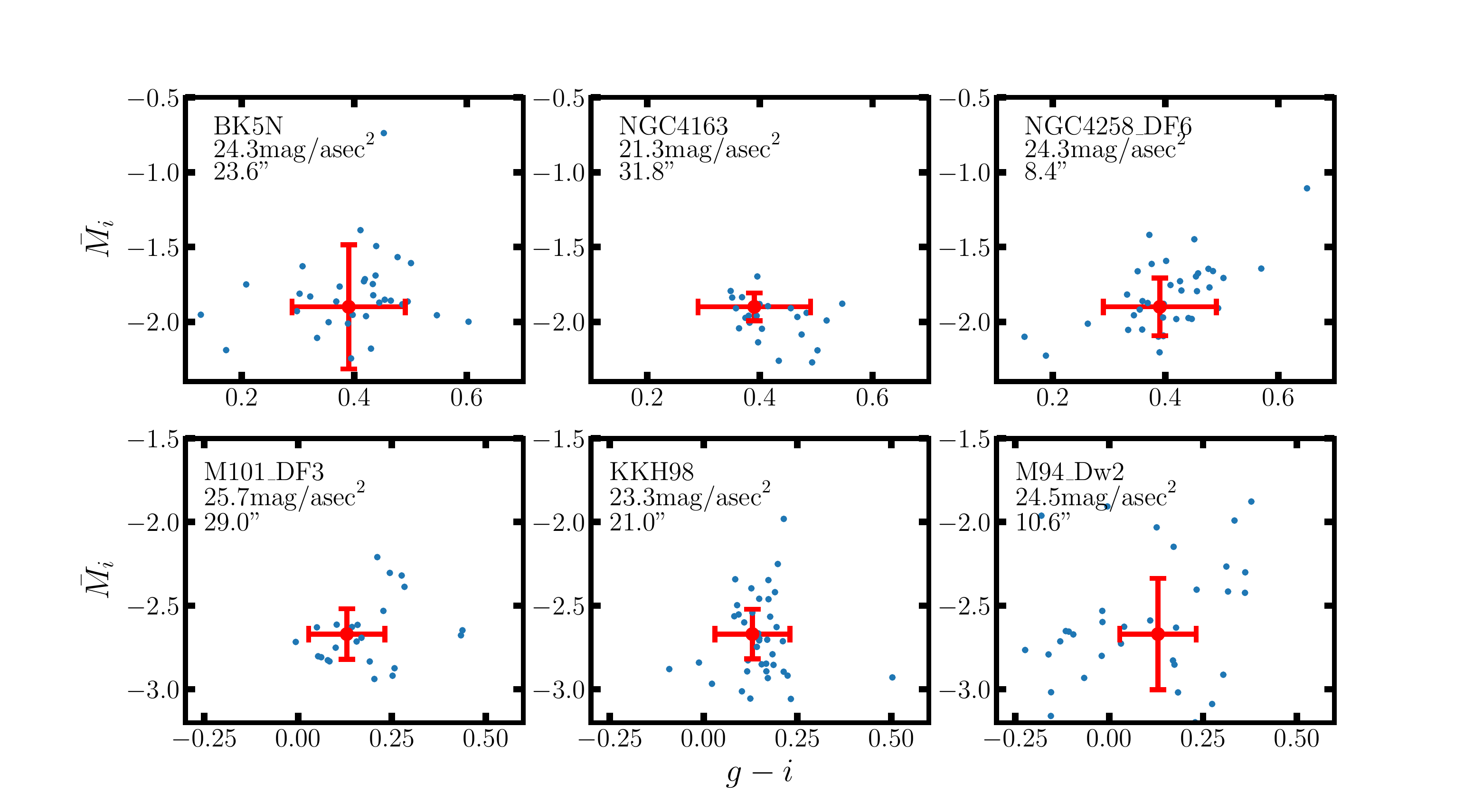}
\caption{Recovery of the color and SBF magnitude in image simulations for six fiducial galaxies in our sample. The red point is at the input color and SBF magnitude of the simulated galaxies and the blue points show the recovered SBF magnitude and color for each simulated galaxy. The vertical errorbars for the red points are the (1$\sigma$) uncertainties calculated for the real galaxy, as described in the text. The horizontal errorbars are 0.1 mag which we take as the contribution to the photometric uncertainty due to the sky subtraction. The galaxy names, central $i$-band surface brightness, and effective radii and listed in the corner of each panel.}
\label{fig:sim_6gals}
\end{figure*}

We do these simulations for six specific galaxies in our sample: BK5N, NGC 4163, NGC 4258\_DF6, M101\_DF3, KKH98, and M94\_Dw2. These six were chosen as they have a representative range in surface brightness, size, and exposure time (cf. Table \ref{tab:sample}). These six galaxies are shown in Figure \ref{fig:app_6gals} in Appendix \ref{app:images}. BK5N, NGC 4163, and NGC 4258\_DF6 were simulated with a [Fe/H]=-1.5 and $\log({\rm age})=9.2$ isochrone and M101\_DF3, KKH98, and M94\_Dw2 were simulated with a [Fe/H]=-0.5 and $\log({\rm age})=8.3$ isochrone. We emphasize that we do not expect that these stellar populations are actually representative of the galaxies (these are likely much younger than the real galaxies) but these isochrones are chosen because they have absolute SBF magnitudes of $\bar{M}_i=-1.90$ mag and $\bar{M}_i=-2.67$ mag, respectively. These SBF magnitudes are roughly representative of the magnitudes we measure in the actual galaxies (see \S\ref{sec:calib}). The older isochrone has color $g-i=0.39$ mag and the younger isochrone has color $g-i=0.13$ mag. These are both on the blue side of our galaxy sample, but this does not affect the suitability of the simulations.

For each of these six galaxies, 50 artificial galaxies are generated and inserted somewhere on the CCD chips near the real galaxy (adjacent chips are also used). The chips are processed in the same way as for the real galaxies with the same sky subtraction, resampling, and coaddition process. Each of the artificial galaxies is then cutout and undergoes the same SBF measurement procedure that the real galaxies do. Simulated galaxies that have fitted S\'{e}rsic parameters that are more than 100\% different from the input S\'{e}rsic parameters are removed. These are commonly due to bright foreground stars making the fits unstable. Generally 30-40 artificial galaxies remain for each of the six galaxies. We can then measure the absolute SBF magnitude and color of each of the simulated galaxies and compare with the input values.

Figure \ref{fig:sim_6gals} shows the results of these simulations for the six fiducial galaxies. The plots show the recovered SBF magnitude and galaxy color of the artificial galaxies compared to the input values. The SBF magnitudes and colors are always concentrated around the inputted values, verifying our measurement methods. The vertical errorbars on the red points in Figure \ref{fig:sim_6gals} are the ($1\sigma$) uncertainties calculated for each (real) galaxy considering the measurement error and uncertainty in the residual variance, as described above. We see that these errorbars appear to realistically represent the spread in the recovered SBF magnitudes. We take this to indicate that our calculated SBF magnitude uncertainties are realistic and include the dominant sources of error. Regarding the color recovery, we see that the recovered colors are generally precise to $\sim0.1$ mag. This appears to be the limit of our sky subtraction procedure. We include this uncertainty in all of our reported colors and magnitudes. The horizontal errorbars in Figure \ref{fig:sim_6gals} show this 0.1 mag uncertainty. The color recovery for M94\_Dw2 appears to be substantially worse than $\pm0.1$ mag. We note that while M94\_Dw2 does not have the lowest surface brightness in our galaxy sample, it does have one of the lowest exposure times and is probably the lowest S/N galaxy in our entire sample. We therefore still take 0.1 mag as a representative uncertainty for the colors of the galaxies due to the sky subtraction.

\section{Details of Data Used}\label{app:data}
In Table \ref{tab:sample2} we give further details of the data used in this project, including P.I.'s, proposal I.D.'s, and the specific filters used.

\begin{deluxetable}{ccccc}
\tablecaption{More information on the main galaxy sample.}
\label{tab:sample2}
\tablehead{\colhead{Name} & \colhead{P.I.} & \colhead{Proposal I.D.} & \colhead{$i$-filter} & \colhead{$g$-filter}}
\startdata
FM1                 & Cuillandre   & 04BF01,05AF17&I.MP9701 & G.MP9701 \\
UGC 004483           & Wilkinson    & 06BO04  &  I.MP9701 & G.MP9701 \\
KDG 061            & Ibata        & 06AF19  &  I.MP9701 & G.MP9701 \\
BK5N                & Cuillandre   & 06AD97   & I.MP9701 & G.MP9701 \\
LVJ1228+4358        & Higgs        & 16AD94 & I.MP9703 & G.MP9702 \\
DDO 125              & Higgs        & 16AD94 & I.MP9703 & G.MP9702 \\
UGCA 365             & Li           & 13AS03 & I.MP9702 & G.MP9701 \\
M94\_dw2             & Li           & 13AS03 & I.MP9702 & G.MP9701 \\
DDO 044              & Davidge      & 03BC03 & I.MP9701 & G.MP9701 \\
NGC 4163             & Higgs        & 16AD94 & I.MP9703 & G.MP9702 \\
NGC 4190             & Higgs        & 16AD94 & I.MP9703 & G.MP9701 \\
KDG090              & Higgs        & 16AD94 & I.MP9703 & G.MP9701 \\
UGC 08508            & Higgs        & 16AD94 & I.MP9703 & G.MP9701 \\
DDO 190              & Waerbeke     & 16AC36 & I.MP9703 & G.MP9701 \\
KKH 98               & McConnachie  & 13AC13,13BC02&I.MP9702 & G.MP9701 \\
Do1                 & Ibata        & 03BF10,05BF48&I.MP9701 & G.MP9701 \\
LVJ1218+4655        & Harris,Ngeow & 10AT01,11AC08&I.MP9702 & G.MP9701 \\
NGC 4258\_DF6         & Harris,Ngeow & 10AT01,11AC08&I.MP9702 & G.MP9701 \\
KDG 101              & Harris,Ngeow & 10AT01,11AC08&I.MP9702 & G.MP9701 \\
M101\_DF1            & CFHTLS       & 05AL02,06AL99&I.MP9701 & G.MP9701 \\
M101\_DF2            & CFHTLS       & 05AL02 & I.MP9701 & G.MP9701 \\
M101\_DF3            & CFHTLS       & 08AL05 & I.MP9702 & G.MP9701 \\
UGC 9405             & CFHTLS       & 07AL02 & I.MP9701 & G.MP9701 \\
M96\_DF9             & Duc          & 09BF07&  I.MP9702 & G.MP9701 \\
M96\_DF1             & Cuillandre Duc Harris & 09AF05 09BF07 10AD94 11AC08&I.MP9702 & G.MP9701 \\
M96\_DF8             & Duc Harris   & 09BF07 11AC08&I.MP9702 & G.MP9701 \\
M96\_DF4             & Duc          & 09AF05 09BF07&I.MP9702 & G.MP9701 \\
M96\_DF5             & Duc Harris   & 09AF05 09BF07 11AC08&I.MP9702 & G.MP9701 \\
M96\_DF7             & Cuillandre Duc Harris & 09BF07 10AD94 11AC08&I.MP9702 & G.MP9701 \\
M96\_DF10            & Duc Harris   & 09BF07 11AC08&I.MP9702 & G.MP9701 \\
M96\_DF6             & Cuillandre Duc Harris & 09AF05 09BF07 10AD94 11AC08&I.MP9702 & G.MP9701 \\
M96\_DF2             & Cuillandre Duc Harris & 09AF05 09BF07 10AD94 11AC08&I.MP9702 & G.MP9701 \\
\enddata
\end{deluxetable}

\section{Different CFHT Filters}\label{sec:app_filters}
In this section, we demonstrate that the differences between the various filter generations used in the CFHT data is insignificant. We focus on the difference between the two $g$ filters (\textsc{G.MP9401} and \textsc{G.MP9402}) and the three $i$ filters (\textsc{I.MP9701}, \textsc{I.MP9702}, and \textsc{I.MP9703}). \footnote{\url{http://www.cfht.hawaii.edu/Instruments/Imaging/MegaPrime/PDFs/megacam.pdf}} and \footnote{\url{http://www.cadc-ccda.hia-iha.nrc-cnrc.gc.ca/en/megapipe/docs/filt.html}} provide filter transforms between the different CFHT filters and to external filter systems. From these relations, \textsc{I.MP9702} differs more from \textsc{I.MP9701} and \textsc{I.MP9703} than those two differ from each other. \textsc{I.MP9701} and \textsc{I.MP9703} differ at most from each other by $\sim$0.004 mag in the color range of our galaxy sample. \textsc{I.MP9703} and \textsc{I.MP9702} differ by $\lesssim0.035$ mag for $r-i\lesssim0.5$ mag which will include our sample. \textsc{G.MP9401} and \textsc{G.MP9402} differ by $\lesssim0.06$ mag for $g-r\lesssim0.7$ mag which is, again, redder than most galaxies in our sample. From these, the effect of the different filters will be fairly insignificant to the color of the measured galaxies. It will be sub-dominant to the 0.1 mag uncertainty that comes from the sky subtraction process and we therefore do not try to correct for the different filters in the galaxy photometry. 

\begin{figure*}
\includegraphics[width=0.95\textwidth]{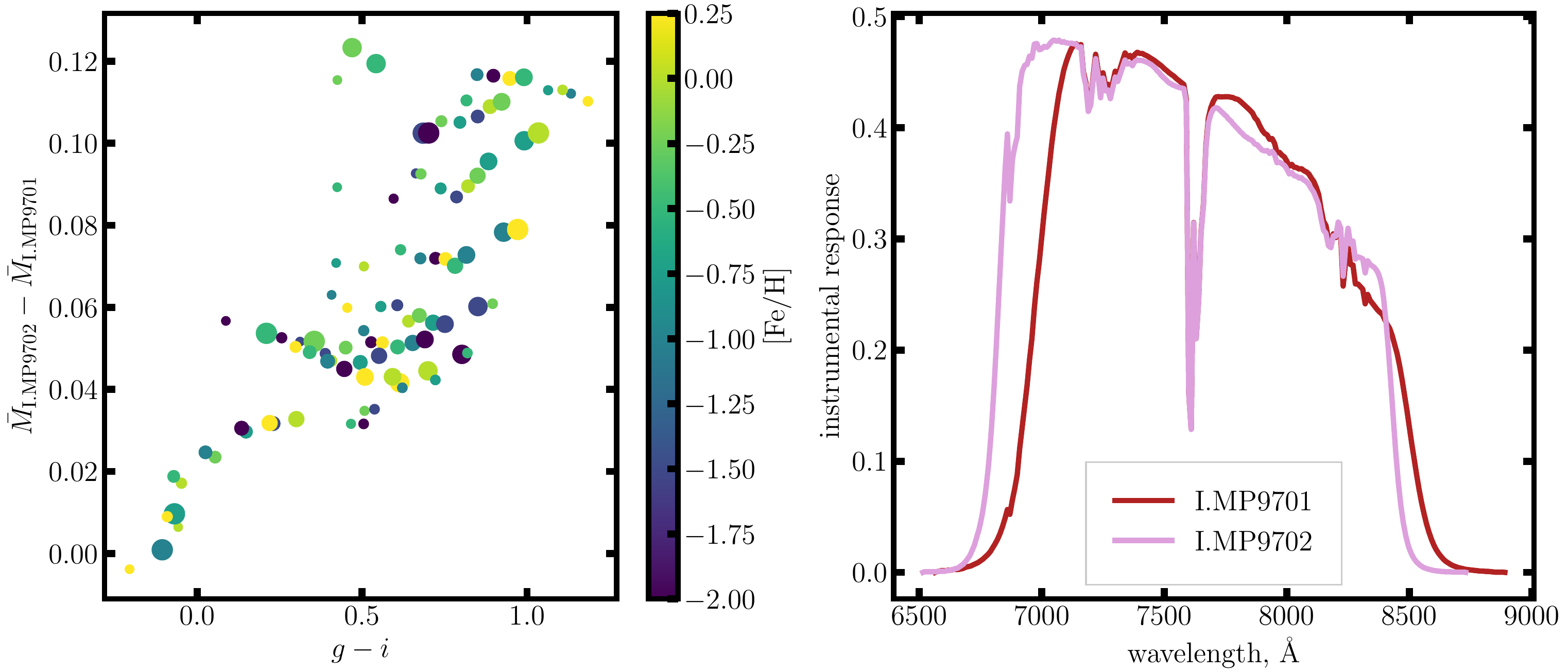}
\caption{\textit{Left}: Difference in the SBF magnitudes in two of the $i$ filters from MIST isochrones versus color for a wide range in metallicities and ages. Point size reflects the age with the biggest points being 13 Gyr populations and the smallest being 1 Gyr populations. \textit{Right}: Difference in in the instrumental response (including atmospheric absorption) for the two $i$ filters.}
\label{fig:filter}
\end{figure*}

To explore the effect of the filter on the SBF magnitude, we use the MIST \citep{mist_models} isochrones which provide luminosity functions for both \textsc{I.MP9701} and \textsc{I.MP9702}. In Figure \ref{fig:filter}, we show the expected difference in SBF magnitude between these two filters as a function of $g-i$ color. For the colors of the galaxies in our sample, the difference is $\lesssim0.06$ mag. We note that the difference between \textsc{I.MP9703} and \textsc{I.MP9702} will be of a similar size. This difference is very sub-dominant to the measurement errors in the SBF measurement and we do not try to correct for this when making the calibration. 

It is possible that the different filters used will contribute to some of the scatter in the calibration but we found that the scatter is $\lesssim0.32$ mag which is much greater than the filter effects which are of the order 0.05 mag.

\section{Rejected Galaxies}\label{sec:gal_rejects}
For the sake of completeness, we list the galaxies with TRGB distances in the catalog of \citet{karachentsev} that were not used in the SBF calibration. Table \ref{tab:gal_rejects} lists these galaxies, along with the primary reason for their exclusion.

\begin{deluxetable}{cc}
\tablecaption{Galaxies rejected from the Calibration Sample.}
\label{tab:gal_rejects}
\tablehead{\colhead{Name} & \colhead{Reason} }
\startdata
And IV & Behind M31's Halo \\ 
BK3N & Too resolved \\ 
Holm IX & Too resolved \\ 
A0952+69 & Irregular \\ 
d0959+68 & Not visible in CFHT data \\ 
NGC 3077 & Significant dust lanes \\ 
Garland & Too resolved \\ 
d1005+68 & Not visible in CFHT data \\ 
MCG+06-27-017 & Non-S\'{e}rsic \\ 
GR34 & Non-S\'{e}rsic \\
NGC 4449 & Irregular \\ 
IC3583 & Irregular \\ 
VCC2037 & Irregular \\ 
dw1335-29  & Not visible in CFHT data \\ 
KK208 & Not visible in CFHT data \\ 
Scl-MM-Dw2 & Not visible in CFHT data \\ 
NGC 404 & In halo of extremely bright star \\ 
KK35 & Too resolved \\ 
UGCA 086 & Poor quality CFHT data \\ 
NGC 4214 &  Non-S\'{e}rsic \\
NGC 253 & Spiral \\ 
NGC 247 & Spiral \\
M81 & Spiral \\ 
M82 & Spiral \\ 
NGC 3384 &   Non-S\'{e}rsic \\
M105 &   Non-S\'{e}rsic \\
M66 &   Non-S\'{e}rsic \\
M104 &   Non-S\'{e}rsic \\
M64 &   Non-S\'{e}rsic \\
NGC 5102 &   Non-S\'{e}rsic \\
Cen A &   Non-S\'{e}rsic \\
M83 & Spiral \\
\enddata
\end{deluxetable}

\section{Images of Select Galaxies}\label{app:images}
\begin{figure*}
\includegraphics[width=0.95\textwidth]{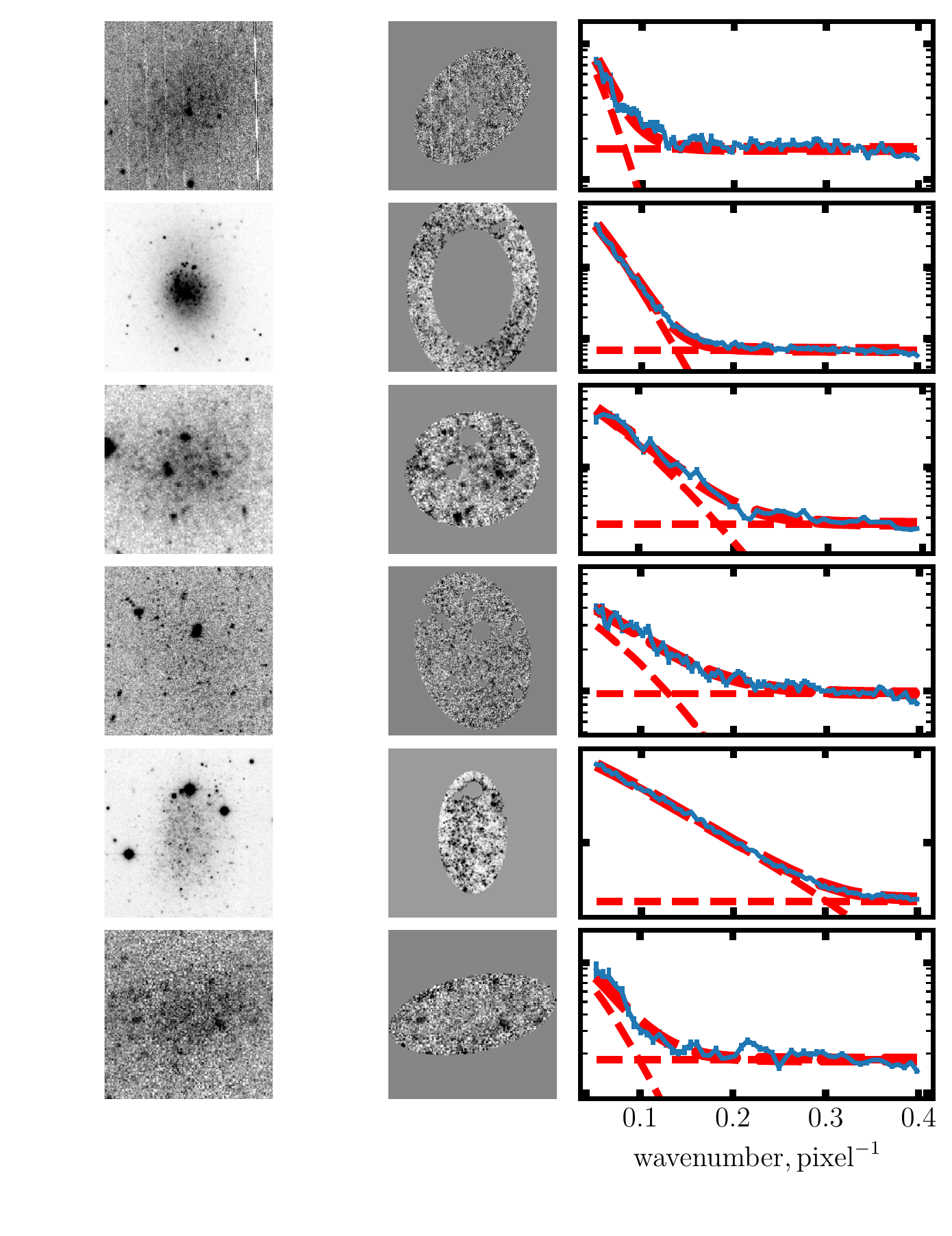}
\caption{Left-most panel shows the $i$ band images for the galaxy, middle panel shows the masked and normalized images, and the right panel shows the fluctuation power spectrum and fit. The galaxies are, in order from top to bottom: BK5N, NGC 4163, NGC4256-DF6, M101-DF3, KKH98, M94-Dw2.}
\label{fig:app_6gals}
\end{figure*}

Here we show $i$ band images, masked and normalized images, and power spectra fits for select galaxies. Figure \ref{fig:app_6gals} shows the six galaxies used in the simulations in the image simulations. 

\end{document}